\documentclass[useAMS,usenatbib]{mn2e}
\usepackage{graphicx}
\usepackage{epsfig}
\usepackage{amssymb}
\usepackage{color}
\usepackage{longtable}
\newcommand{\mincir}{\raise
-3.truept\hbox{\rlap{\hbox{$\sim$}}\raise4.truept\hbox{$<$}\ }}
\newcommand{\magcir}{\raise
-3.truept\hbox{\rlap{\hbox{$\sim$}}\raise4.truept\hbox{$>$}\ }}
\newcommand{\minmag}{\raise
-3.truept\hbox{\rlap{\hbox{$<$}}\raise5.truept\hbox{$<$}\ }}
\newcommand{\beqa}{\begin{eqnarray}}
\newcommand{\eeqa}{\end{eqnarray}}
\newcommand{\be}{\begin{equation}}
\newcommand{\ee}{\end{equation}}
 \newcommand{\ba}{\begin{eqnarray}}
\newcommand{\ea}{\end{eqnarray}}
\newcommand{\brr}{\begin{array}}
\newcommand{\err}{\end{array}}
\newcommand{\bc}{\begin{center}}
\newcommand{\ec}{\end{center}}

\newcommand{\omm}{{\Omega_{\rm m}}}
\newcommand{\epkk}{{E_{\rm peak}}}
\newcommand{\tlag}{{\tau_{\rm lag}}}
\newcommand{\trt}{{\tau_{RT}}}

\newcommand{\pbo}{{P_{\rm bolo}}}
\newcommand{\sbo}{{S_{\rm bolo}}}

\title[The updated luminosity correlations of gamma-ray bursts and cosmological implications]
{The updated luminosity correlations of gamma-ray bursts and
cosmological implications}
\author[Fa-Yin Wang, Shi Qi \& Zi-Gao Dai]{Fa-Yin Wang$^{1,2}$\thanks{fayinwang@nju.edu.cn},
Shi Qi$^{3,4,5}$\thanks{qishi11@gmail.com} \& Zi-Gao Dai$^{1,2}$\thanks{dzg@nju.edu.cn}\\
\vspace{0.1cm} $^1$Department of Astronomy, Nanjing University,
Nanjing 210093, China\\
$^2$Key Laboratory of Modern Astronomy and Astrophysics (Nanjing
University), Ministry of Education, Nanjing 210093, China\\
$^3$Purple Mountain Observatory, Chinese Academy of Sciences,
Nanjing 210008, China \\
$^4$Joint Center for Particle, Nuclear Physics and Cosmology,
Nanjing University - Purple Mountain Observatory, Nanjing 210093,
China \\
$^5$Key Laboratory of Dark Matter and Space Astronomy, Chinese Academy
of Sciences.
}

\begin{document}

\maketitle

\begin{abstract}
Several interesting luminosity correlations among gamma-ray burst
(GRB) variables have been recently discussed extensively. In this
paper, we derive the six luminosity correlations
($\tlag-L$, $V-L$, $\epkk-L$, $\epkk-E_\gamma$, $\trt-L$,
$\epkk-E_{\gamma, \mathrm{iso}}$) from the
light curves and spectra of the latest 116 long GRBs, including the
time lag ($\tlag$) between low and high photon energy light curves,
the variability ($V$) of the light curve, the peak energy of the
spectrum ($\epkk$), and the minimum rise time ($\trt$) of the peaks.
We find that the intrinsic scatter of the $V-L$ correlation is too
large and there seems no inherent correlation between the two
parameters using the latest GRB data. The other five correlations
indeed exist when the sample is enlarged. The $\epkk-E_\gamma$
correlation has a significantly lower intrinsic scatter compared
to the other correlations. We divide the full data into four
redshift bins when testing possible evolution of the correlations
with redshift. We find no statistically significant evidence for the
redshift evolution of the luminosity correlations.
To avoid the circularity problem when constraining the
cosmological parameters, we simultaneously minimize $\chi^2$ with
respect to both correlation parameters $a$, $b$ and the cosmological
parameters using the maximum likelihood method. For the flat
$\Lambda$CDM, the best fit is $\omm=0.31^{+0.13}_{-0.10}$. We also
constrain the possible evolution of the equation of state (EOS) of
the dark energy using the GRBs together with the Union2 compilation
of SNe Ia and the $H(z)$ data. The result is consistent with the
cosmological constant at $2 \sigma$ confidence level and mainly due
to the GRB data, the dark energy EOS shows slight deviation from
$-1$ at $z \geq0.5$ as was persistently presented with many previous
data sets.

\end{abstract}


\vspace{1.0cm}

\section{Introduction}
Unexpected accelerating expansion of the universe was first
discovered by observing type Ia supernovae (SNe Ia) (Riess et al.
1998; Perlmutter et al. 1999). Independent observations from
baryonic acoustic oscillations (BAO) (Eisenstein et al. 2005;
Percival et al. 2007), the anisotropy spectrum of cosmic microwave
background radiation (Komatsu et al. 2009) and the large scale
structure data from large galaxy redshift surveys (Tegmark et al.
2006) have confirmed this surprising result. This acceleration is
commonly attributed to dark energy, which is the most mysterious
problem in modern cosmology. Among parameters that describe the
properties of dark energy, the equation of state (EOS) is one of the
most important. Whether and how it evolves with time is crucial in
distinguishing different cosmological models. A nearly model-independent
approach in which uncorrelated estimates are made about discrete $w(z)$
at different redshifts has been extensively discussed (Huterer \&
Cooray 2005; Riess et al. 2007; Sullivan et al. 2007; Qi, Wang \& Lu
2008).

In order to measure the expansion history of our Universe, we need
the Hubble diagram of standard candles. SNe Ia  are the well known
standard candles that have played an important role in constraining
cosmological parameters. Unfortunately, it is difficult to observe
SNe Ia at $z>1.7$, even with excellent space based projects such as
SNAP (Aldering et al. 2004). They cannot provide any information on
the cosmic expansion beyond redshift 1.7. With gamma-ray bursts
(GRBs), we can access much higher redshifts. The high luminosities
of GRBs make them detectable out to the edge of the visible universe
(Lamb \& Reichart 2000; Ciardi \& Loeb 2000; Bromm \& Loeb 2002,
2006). The farthest GRB observed hitherto is GRB 090423 at $z=8.2$
(Tanvir et al. 2009; Salvaterra et al. 2009). Schaefer (2007)
complied 69 GRBs to make simultaneous uses of five luminosity
relations, which are the correlations of $\tau_{\rm lag}-L$ (Norris,
Marani \& Bonnell 2000), $V-L$ (Fenimore \& Ramirez-Ruiz 2000),
$E_{\rm peak}-L$ (Schaefer et al. 2003; Wei \& Gao 2003), $E_{\rm
peak}-E_{\gamma}$(Ghirlanda et al. 2004a), and $\tau_{\rm RT}-L$
(Schaefer 2007). Here the time lag ($\tau_{\rm lag}$) is the time
shift between the hard and soft light curves, $L$ is the peak
luminosity of a GRB, the variability $V$ of a burst denotes whether
its light curve is spiky or smooth and it can be obtained by
calculating the normalized variance of an observed light curve
around a smoothed version of that light curve (Fenimore \& Ramirez-
Ruiz 2000), $E_{\rm peak}$ is the photon energy at which the $\nu
F_{\nu}$ spectrum peaks, $E_{\gamma}=(1-\cos\theta_j)E_{\rm \gamma,
iso}$ is the collimation-corrected energy of a GRB, and the minimum
rise time ($\tau_{\rm RT}$) in the gamma-ray light curve is the
shortest time over which the light curve rises by half of the peak
flux of the pulse. More recently, Yu et al. (2009) found that, for
the three-dimensional (3D) luminosity relations between the
luminosity and an energy scale $\epkk$ and a timescale ($\tlag$ or
$\trt$), the intrinsic scatters are considerably smaller than those
of corresponding two-dimensional (2D) luminosity relations. Dainotti
et al. (2008, 2010) and Qi \& Lu (2010) found new correlations
between the transition times of the X-ray light curve from exponential
to power law and the X-ray luminosities at the transitions.
After being calibrated with luminosity
relations, GRBs may be used as standard candles to provide
information on cosmic expansion at high redshifts and, at the same
time, to tighten the constraints on cosmic expansion at low
redshifts (Dai et al. 2004; Ghirlanda et al. 2004b; Friedman \&
Bloom 2005; Liang \& Zhang 2005, 2006; Wang \& Dai 2006; Schaefer
2007; Wright 2007; Wang, Dai \& Zhu 2007; Wang 2008; Qi, Wang \& Lu
2008a,b; Liang et al. 2008; Amati et al. 2008; Cardone et al. 2009,
2010; Liang et al. 2009; Qi, Lu \& Wang 2009; Izzo et al. 2009;
Liang \& Zhu 2010). GRBs also can potentially probe the cosmographic
parameters to distinguish between dark energy and modified gravity
models (Wang, Dai  \& Qi 2009a, b; Vitagliano et al. 2010;
Capozziello \& Izzo 2008).

The correlations among GRB variables span a very large range in
redshift. Possible evolution effect must be considered when we use
these correlations. Li (2007) used the Amati relation ($E_{\rm
peak}-E_{\gamma, \mathrm{iso}}$) (Amati et al. 2002) as an example
to test the cosmic evolution of GRBs and found that the slope of the
correlation evolves with the redshift. In contrast, Basilakos \&
Perivolaropoulos (2008) found no statistically significant evidence
for redshift dependence of correlation slopes using 69 GRBs. In this
paper, we first enlarge the GRB sample with the new data from Xiao
\& Schaefer (2009). Our sample includes 116 GRBs ranging from
$z=0.17$ to $z=8.2$. We divide these GRBs into four redshift bins to
investigate the possible evolution effect. Here the focus is on the
correlations, so we fix the cosmological parameters. We also use
GRBs to constrain the cosmological parameters and dark energy EOS.
In order to avoid the circularity problem, we simultaneously fit the
correlation parameters and the cosmological parameters.

The structure of this paper is as follows: in the next section we
show the latest GRB data and describe our fitting methods. In
section 3 we present the updated luminosity correlations and test
their redshift dependence. Constraints on cosmological parameters
and equation of state of dark energy are presented in section 4.
Some conclusions are presented in section 5.

\section{Observational data and analysis method}

The luminosity correlations we will discuss here typically relate a
GRB observable with the isotropic peak luminosity $L$ (it is also
referenced to as $L_p$ in many papers), the isotropic
energy $E_{\rm \gamma, iso}$, or the collimation-corrected energy
$E_{\gamma}$. The isotropic peak luminosity is given by
\be
 L = 4\pi d^2_{L}P_{\rm bolo}, \label{ldl}
\ee
the isotropic energy is
\be
E_{\rm \gamma, iso} = 4\pi d^2_{L}S_{\rm bolo}(1+z)^{-1},
\ee
and the collimation-corrected energy is
\be
E_{\gamma}=E_{\rm \gamma, iso}F_{\rm beam}
=4\pi d^2_{L}S_{\rm bolo}F_{\rm beam}(1+z)^{-1}.
\label{egdl}
\ee
Here, $P_{\rm bolo}$ and
$S_{\rm bolo}$ are the bolometric peak flux and fluence,
respectively, while $F_{\rm beam} = 1 - \cos{\theta_{\rm jet}}$ is
the beaming factor. From Sari, Piran, \& Halpern (1999),
\begin{equation}
  \theta_{\rm jet} = 0.161 [t_{\rm jet}/(1+z)]^{3/8}
  (n~\eta _{\gamma} / E_{\rm \gamma,iso,52})^{1/8},
\end{equation}
where $z$ is the redshift, $t_{\rm jet}$ is the jet break time
measured in days, $n$ is the density of the circumburst medium in
particles per cubic centimeter, $\eta _{\gamma}$ is the radiative
efficiency, and $E_{\rm \gamma,iso,52}$ is the isotropic energy in units of
$10^{52}$ erg for an Earth-facing jet. The jet break time ($t_{\rm
jet}$) can be measured when the afterglow brightness has a power-law
decline that suddenly steepens due to the slowing down of the jet
until the relativistic beaming roughly equals the jet opening angle.
In the absence of these detailed fits, we adopt $\eta_\gamma=0.2$
and $n=3~$cm$^{-3}$ (Schaefer 2007). Note that $P_{\rm bolo}$ and
$S_{bolo}$ are computed from the observed GRB energy spectrum
$\Phi(E)$ as follows (Ghirlanda et al. 2004a, Amati 2006):
\begin{equation}
P_{\rm bolo} = P  \ {\times} \ \frac{\int_{1/(1 + z)}^{10^4/(1 +
z)}{E \Phi(E) dE}} {\int_{E_{\rm min}}^{E_{\rm max}}{\Phi(E) dE}} \
, \label{eq: defpbolo}
\end{equation}
\begin{equation}
S_{\rm bolo} = S \ {\times} \ \frac{\int_{1/(1 + z)}^{10^4/(1 +
z)}{E \Phi(E) dE}} {\int_{E_{\rm min}}^{E_{\rm max}}{E \Phi(E) dE}}
\ , \label{eq: defsbolo}
\end{equation}
with $P$ and $S$ being the observed peak energy and fluence in units
of ${\rm photons/cm^2/s}$ and ${\rm erg/cm^2}$, respectively, and
$(E_{\rm min}, E_{\rm max})$ the detection thresholds of the
observing instrument. For pre-\emph{Swift} GRBs, we take the values of
$P_{\rm bolo}$ and $S_{\rm bolo}$ directly from Schaefer (2007). For
those GRBs observed by \emph{Swift}, we adopt the values of $P$ and
$S$ from \emph{Swift} website~\footnote{See $\rm
  http://swift.gsfc.nasa.gov/docs/swift/archive/grb\_table.$}
and calculate $P_{\rm bolo}$ and $S_{\rm bolo}$ using the above
formulae. Concerning the errors of $P_{\rm bolo}$ and $S_{\rm bolo}$
during the calculation, we only take into account the errors
propagating from that of $P$ and $S$. The uncertainties from $\Phi(E)$
are absorbed into intrinsic scatters of the correlations.
Note that the energy spectrum is modeled using a smoothly broken
power\,-\,law (Band et al. 1993),
\begin{equation}
\Phi(E) = \left \{
\begin{array}{ll}
A E^{\alpha} {\rm e}^{-(2 + \alpha) E/E_{\rm peak}} & E \le
\frac{\alpha
-\beta}{2 + \alpha}E_{\rm peak} \\ ~ & ~ \\
B E^{\beta} & {\rm otherwise}
\end{array}
\right . \  \label{eq: band}
\end{equation}
where $\alpha$ is the asymptotic power-law index for photon energies
below the break and $\beta$ is the power-law index for photon
energies above the break. We use the values of $\alpha$ and $\beta$
from Xiao \& Schaefer (2009). The luminosity correlations are
power-law relations of either $L$, $E_{\rm \gamma, iso}$ or
$E_{\gamma}$ as a function of $\tlag$, $V$, $E_{\rm peak}$, or $\trt$.
The luminosity indicators of $\tlag$, $V$, $E_{\rm peak}$, and $\trt$
are also directly taken from Xiao \& Schaefer (2009).
$L$, $E_{\rm \gamma, iso}$, and $E_{\gamma}$ depend not only on the
GRB observables $\pbo$ or $\sbo$, but also on the cosmological
parameters through the
luminosity distance $d_L$, which in a flat universe is expressed in
terms of the Hubble expansion rate $H(z)=H_0 E(z)$ as
\begin{equation}
  d_L(\omm,z) = (1+z)\frac{c}{H_0} \int_0^z \frac{dz'}{E(z')},
  \label{dlum1}
\end{equation}
where $E^2(z) = \omm (1+z)^3 + \Omega_{\rm x}f_{\rm x}(z)$ and the
dimensionless dark energy density $f_{\rm x}(z)$ is given by ($w(z)$
is the EOS of dark energy)
\begin{eqnarray}
  \label{eq:fz}
  f_{\rm x}(z)=\exp \left[
    3\int_0^z\frac{1+w(\tilde{z})}{1+\tilde{z}}\mathrm{d}\tilde{z}
  \right]
  .
\end{eqnarray}
When the focus is on the luminosity correlations themselves, the
cosmological parameters here are fixed.

The luminosity correlations involved in this paper are
\begin{eqnarray}
  \label{eq:GRB-lag-L}
  \log \frac{L}{1 \; \mathrm{erg} \; \mathrm{s}^{-1}}
  &=& a_1+b_1 \log
  \left[
    \frac{\tau_{\mathrm{lag}}(1+z)^{-1}}{0.1\;\mathrm{s}}
  \right]
  ,
  \\
  \label{eq:GRB-V-L}
  \log \frac{L}{1 \; \mathrm{erg} \; \mathrm{s}^{-1}}
  &=& a_2+b_2 \log
  \left[
    \frac{V(1+z)}{0.02}
  \right]
  ,
  \\
  \label{eq:GRB-E_peak-L}
  \log \frac{L}{1 \; \mathrm{erg} \; \mathrm{s}^{-1}}
  &=& a_3+b_3 \log
  \left[
    \frac{E_{\mathrm{peak}}(1+z)}{300\;\mathrm{keV}}
  \right]
  ,
  \\
  \label{eq:GRB-E_peak-E_gamma}
  \log \frac{E_{\gamma}}{1\;\mathrm{erg}}
  &=& a_4+b_4 \log
  \left[
    \frac{E_{\mathrm{peak}}(1+z)}{300\;\mathrm{keV}}
  \right]
  ,
  \\
  \label{eq:GRB-tau_RT-L}
  \log \frac{L}{1 \; \mathrm{erg} \; \mathrm{s}^{-1}}
  &=& a_5+b_5 \log
  \left[
    \frac{\tau_{\mathrm{RT}}(1+z)^{-1}}{0.1\;\mathrm{s}}
  \right]
  ,
  \\
 \label{eq:GRB-E_peak-E_iso}
  \log \frac{E_{\gamma,\mathrm{iso}}}{1\;\mathrm{erg}}
  &=& a_6+b_6 \log
  \left[
    \frac{E_{\mathrm{peak}}(1+z)}{300\;\mathrm{keV}}
  \right]
  .
\end{eqnarray}
Concerning the luminosity indicators in the correlations, for the
temporal indicators, the observed quantities must be divided
by $1+z$ to correct the time dilation. The observed $V$-value must
be multiplied by $1+z$ because it varies inversely with time, and
the observed $\epkk$ must be multiplied by $1+z$ to correct the
redshift dilation of the spectrum.

The first five of the correlations listed above were the ones
considered in Schaefer (2007) and Xiao \& Schaefer (2009). We add in
our analysis the investigation of the Amati correlation
($E_{\mathrm{peak}}-E_{\gamma,\mathrm{iso}}$), which was initially
discovered on a small sample of BeppoSAX GRBs with known redshift
(Amati et al. 2002) and confirmed afterwards by Swift observations
(Amati 2006).
Compared to the $E_{\mathrm{peak}}-E_{\gamma}$ correlation, due to the
independence of $\theta_{\mathrm{jet}}$, the
$E_{\mathrm{peak}}-E_{\gamma,\mathrm{iso}}$ correlation can be used
for almost the whole GRB sample and does not suffer from the
assumptions and uncertainties around $\theta_{\mathrm{jet}}$ that
affect the $E_{\mathrm{peak}}-E_{\gamma}$ correlation.
Also, compared to the $E_{\mathrm{peak}}-L$ correlation, the
$E_{\mathrm{peak}}-E_{\gamma,\mathrm{iso}}$ correlation is not
affected by assumptions on the peak flux time scale and on the
spectral shape at the peak (i.e., the peak luminosity is always
computed by assuming the spectral shape of the time-averaged spectrum,
which is not physical, given that the spectrum at the peak is often
much different than the average one).
There is also debate about the reality of these correlations, see
Nakar \& Piran (2005), Band \& Preece 2005, Butler et al. (2007),
Butler et al. (2009), Ghirlanda et
al. (2005), Bosnjak et al. (2008), Ghirlanda et al. (2008), Nava et
al. (2008), Krimm et al. (2009), Amati et al. (2009), Ghirlanda et
al. (2010) etc.

In Table 1, we list the variables
of 116 GRBs that we use in fitting luminosity correlations. In
addition to the GRBs included in the analysis of Xiao \& Schaefer
(2009), we add the GRB090423, which has the highest redshift so
far. We use the typical spectral index $\alpha=-1$ and $\beta=-2.2$
for this burst (Salvaterra et al. 2009).

\onecolumn
{\small
\begin{longtable}{@{} l @{ } c @{\quad} c @{\quad} c @{ } c @{ } c @{ }
    c @{ } c @{ } c @{}}
  \hline
  {\bf GRB} & z & $P_{\rm bolo}$ & $S_{\rm bolo}$ & $F_{\rm beam}$
  & $\tau_{lag}$ & $V$ & $E_{\rm peak}$ & $\tau_{RT}$ \\
  &  &[${\rm erg/cm^{2}s}$] & [${\rm erg/cm^{2}}$] &
  & [sec] & & [keV] &[sec] \\
  \hline
  \endhead
970228 & $0.70$ & 7.3E-6 $\pm$ 4.3E-7 & $\cdots$ & $\cdots$ & $\cdots$ & $0.016 \pm 0.010$ & $115^{+38}_{-38}$ & $\cdots$ \\
970508 & $0.84$ & 3.3E-6 $\pm$ 3.3E-7 & 8.09E-6 $\pm$ 8.1E-7 & 0.0795 $\pm$ 0.0204 & $0.49 \pm 0.02$ & $0.018 \pm 0.004$ & $389^{+40}_{-40}$ & $0.65 \pm 0.07$ \\
970828 & $0.96$ & 1.0E-5 $\pm$ 1.1E-6 & 1.23E-4 $\pm$ 1.2E-5 & 5.32E-03 $\pm$ 1.44E-03 & $\cdots$ & $0.052 \pm 0.005$ & $298^{+30}_{-30}$ & $0.36 \pm 0.14$ \\
971214 & $3.42$ & 7.5E-7 $\pm$ 2.4E-8 & $\cdots$ & $\cdots$ & $0.03 \pm 0.05$ & $0.048 \pm 0.002$ & $190^{+20}_{-20}$ & $\cdots$ \\
980703 & $0.97$ & 1.2E-6 $\pm$ 3.6E-8 & 2.83E-5 $\pm$ 2.9E-6 & 1.84E-02 $\pm$ 2.67E-03 & $0.69 \pm 0.02$ & $0.024 \pm 0.001$ & $254^{+25}_{-25}$ & $3.00 \pm 0.19$ \\
990123 & $1.61$ & 1.3E-5 $\pm$ 5.0E-7 & 3.11E-4 $\pm$ 3.1E-5 & 2.41E-03 $\pm$ 6.90E-04 & $0.07 \pm 0.01$ & $0.059 \pm 0.003$ & $604^{+60}_{-60}$ & $\cdots$ \\
990506 & $1.31$ & 1.1E-5 $\pm$ 1.5E-7 & $\cdots$ & $\cdots$ & $0.04 \pm 0.01$ & $0.337 \pm 0.001$ & $283^{+30}_{-30}$ & $0.13 \pm 0.01$ \\
990510 & $1.62$ & 3.3E-6 $\pm$ 1.2E-7 & 2.85E-5 $\pm$ 2.9E-6 & 2.13E-03 $\pm$ 3.19E-04 & $0.03 \pm 0.01$ & $0.118 \pm 0.001$ & $126^{+10}_{-10}$ & $0.13 \pm 0.01$ \\
990705 & $0.84$ & 6.6E-6 $\pm$ 2.6E-7 & 1.34E-4 $\pm$ 1.5E-5 & 3.48E-03 $\pm$ 9.60E-04 & $\cdots$ & $0.097 \pm 0.004$ & $189^{+15}_{-15}$ & $0.62 \pm 0.37$ \\
991208 & $0.71$ & 2.1E-5 $\pm$ 2.1E-6 & $\cdots$ & $\cdots$ & $\cdots$ & $0.023 \pm 0.003$ & $190^{+20}_{-20}$ & $0.27 \pm 0.01$ \\
991216 & $1.02$ & 4.1E-5 $\pm$ 3.8E-7 & 2.48E-4 $\pm$ 2.5E-5 & 3.00E-03 $\pm$ 9.46E-04 & $0.03 \pm 0.01$ & $0.062 \pm 0.003$ & $318^{+30}_{-30}$ & $0.09 \pm 0.01$ \\
000131 & $4.50$ & 7.3E-7 $\pm$ 8.3E-8 & $\cdots$ & $\cdots$ & $\cdots$ & $0.056 \pm 0.005$ & $163^{+13}_{-13}$ & $0.84 \pm 0.39$ \\
000210 & $0.85$ & 2.0E-5 $\pm$ 2.1E-6 & $\cdots$ & $\cdots$ & $\cdots$ & $0.018 \pm 0.002$ & $408^{+14}_{-14}$ & $0.45 \pm 0.03$ \\
000911 & $1.06$ & 1.9E-5 $\pm$ 1.9E-6 & $\cdots$ & $\cdots$ & $\cdots$ & $0.122 \pm 0.013$ & $986^{+100}_{-100}$ & $0.07 \pm 0.22$ \\
000926 & $2.07$ & 2.9E-6 $\pm$ 2.9E-7 & $\cdots$ & $\cdots$ & $\cdots$ & $0.326 \pm 0.034$ & $100^{+7}_{-7}$ & $\cdots$ \\
010222 & $1.48$ & 2.3E-5 $\pm$ 7.2E-7 & 2.45E-4 $\pm$ 9.1E-6 & 0.0014 $\pm$ 0.0001 & $\cdots$ & $0.143 \pm 0.004$ & $309^{+12}_{-12}$ & $0.45 \pm 0.01$ \\
010921 & $0.45$ & 1.8E-6 $\pm$ 1.6E-7 & $\cdots$ & $\cdots$ & $1.00 \pm 0.04$ & $0.008 \pm 0.006$ & $89^{+22}_{-14}$ & $4.31 \pm 0.71$ \\
020124 & $3.20$ & 6.1E-7 $\pm$ 1.0E-7 & 1.14E-5 $\pm$ 1.1E-6 & 4.10E-03 $\pm$ 1.09E-03 & $0.07 \pm 0.06$ & $0.266 \pm 0.040$ & $87^{+18}_{-12}$ & $0.59 \pm 0.17$ \\
020405 & $0.70$ & 7.4E-6 $\pm$ 3.1E-7 & 1.10E-4 $\pm$ 2.1E-6 & 5.98E-03 $\pm$ 1.96E-03 & $\cdots$ & $0.104 \pm 0.007$ & $364^{+90}_{-90}$ & $0.48 \pm 0.09$ \\
020813 & $1.25$ & 3.8E-6 $\pm$ 2.6E-7 & 1.59E-4 $\pm$ 2.9E-6 & 1.14E-03 $\pm$ 2.92E-04 & $0.15 \pm 0.01$ & $0.164 \pm 0.004$ & $140^{+14}_{-13}$ & $0.59 \pm 0.05$ \\
021004 & $2.32$ & 2.3E-7 $\pm$ 5.5E-8 & 3.61E-6 $\pm$ 8.6E-7 & 1.04E-02 $\pm$ 2.56E-03 & $0.71 \pm 0.19$ & $0.035 \pm 0.067$ & $80^{+53}_{-23}$ & $1.23 \pm 0.96$ \\
021211 & $1.01$ & 2.3E-6 $\pm$ 1.7E-7 & $\cdots$ & $\cdots$ & $0.31 \pm 0.01$ & $0.006 \pm 0.003$ & $46^{+8}_{-6}$ & $0.57 \pm 0.01$ \\
030115 & $2.50$ & 3.2E-7 $\pm$ 5.1E-8 & $\cdots$ & $\cdots$ & $0.44 \pm 0.06$ & $0.020 \pm 0.020$ & $83^{+53}_{-22}$ & $0.70 \pm 0.40$ \\
030226 & $1.98$ & 2.6E-7 $\pm$ 4.7E-8 & 8.33E-6 $\pm$ 9.8E-7 & 2.72E-03 $\pm$ 6.82E-04 & $0.31 \pm 0.22$ & $0.033 \pm 0.029$ & $97^{+27}_{-17}$ & $1.76 \pm 1.15$ \\
030323 & $3.37$ & 1.2E-7 $\pm$ 6.0E-8 & $\cdots$ & $\cdots$ & $\cdots$ & $0.021 \pm 0.338$ & $44^{+90}_{-26}$ & $\cdots$ \\
030328 & $1.52$ & 1.6E-6 $\pm$ 1.1E-7 & 6.14E-5 $\pm$ 2.4E-6 & 1.96E-03 $\pm$ 4.92E-04 & $0.08 \pm 0.08$ & $0.024 \pm 0.003$ & $130^{+14}_{-13}$ & $1.69 \pm 0.81$ \\
030329 & $0.17$ & 2.0E-5 $\pm$ 1.0E-6 & 2.31E-4 $\pm$ 2.0E-6 & 4.89E-03 $\pm$ 8.62E-04 & $0.15 \pm 0.01$ & $0.065 \pm 0.002$ & $68^{+2}_{-2}$ & $0.66 \pm 0.01$ \\
030429 & $2.66$ & 2.0E-7 $\pm$ 5.4E-8 & 1.13E-6 $\pm$ 1.9E-7 & 5.76E-03 $\pm$ 2.79E-03 & $0.03 \pm 0.17$ & $0.220 \pm 0.135$ & $35^{+12}_{-8}$ & $\cdots$ \\
030528 & $0.78$ & 1.6E-7 $\pm$ 3.2E-8 & $\cdots$ & $\cdots$ & $12.56 \pm 0.14$ & $0.017 \pm 0.010$ & $32^{+5}_{-5}$ & $2.13 \pm 0.42$ \\
040924 & $0.86$ & 2.6E-6 $\pm$ 2.8E-7 & $\cdots$ & $\cdots$ & $0.90 \pm 0.01$ & $0.060 \pm 0.003$ & $67^{+6}_{-6}$ & $0.33 \pm 0.17$ \\
041006 & $0.71$ & 2.5E-6 $\pm$ 1.4E-7 & 1.75E-5 $\pm$ 1.8E-6 & 1.13E-03 $\pm$ 3.40E-04 & $\cdots$ & $0.050 \pm 0.002$ & $63^{+13}_{-13}$ & $1.28 \pm 0.01$ \\
050126 & $1.29$ & 1.07E-07 $\pm$ 1.56E-08 & 1.99E-06 $\pm$ 1.15E-07 & $\cdots$ & $2.74 \pm 0.02$ & $-0.010 \pm 0.065$ & $47^{+23}_{-8}$ & $1.58 \pm 1.91$ \\
050223 & $0.59$ & 1.18E-07 $\pm$ 1.66E-08 & 1.68E-06 $\pm$ 1.04E-07 & $\cdots$ & $\cdots$ & $0.111 \pm 0.094$ & $62^{+10}_{-10}$ & $\cdots$ \\
050315 & $1.95$ & 2.79E-07 $\pm$ 1.93E-08 & 7.52E-06 $\pm$ 2.07E-07 & $\cdots$ & $\cdots$ & $0.032 \pm 0.016$ & $39^{+7}_{-7}$ & $1.97 \pm 1.62$ \\
050401 & $2.90$ & 1.74E-06 $\pm$ 9.09E-08 & 1.69E-05 $\pm$ 3.83E-07 & 2.20E-03 $\pm$ 7.52E-04 & $0.06 \pm 0.02$ & $0.187 \pm 0.019$ & $118^{+18}_{-18}$ & $0.25 \pm 0.16$ \\
050406 & $2.44$ & 4.05E-08 $\pm$ 6.84E-09 & 1.41E-07 $\pm$ 1.77E-08 & $\cdots$ & $\cdots$ & $0.020 \pm 0.274$ & $25^{+35}_{-13}$ & $\cdots$ \\
050408 & $1.24$ & 1.1E-6 $\pm$ 2.1E-7 & $\cdots$ & $\cdots$ & $0.31 \pm 0.02$ & $0.082 \pm 0.005$ & $100^{+100}_{-50}$ & $0.49 \pm 0.02$ \\
050416A & $0.65$ & 5.41E-07 $\pm$ 3.24E-08 & 9.28E-07 $\pm$ 5.68E-08 & 1.45E-02 $\pm$ 8.38E-03 & $\cdots$ & $0.021 \pm 0.030$ & $15^{+2}_{-3}$ & $0.54 \pm 0.06$ \\
050505 & $4.27$ & 2.94E-07 $\pm$ 2.99E-08 & 5.23E-06 $\pm$ 2.29E-07 & $\cdots$ & $0.71 \pm 0.13$ & $0.076 \pm 0.031$ & $70^{+140}_{-24}$ & $0.60 \pm 0.21$ \\
050525A & $0.61$ & 4.74E-06 $\pm$ 6.50E-08 & 2.44E-05 $\pm$ 2.14E-07 & 2.47E-03 $\pm$ 8.46E-04 & $0.12 \pm 0.01$ & $0.093 \pm 0.003$ & $81^{+1}_{-1}$ & $0.32 \pm 0.01$ \\
050603 & $2.82$ & 8.01E-06 $\pm$ 2.42E-07 & 2.73E-05 $\pm$ 5.98E-07 & $\cdots$ & $-0.01 \pm 0.01$ & $0.125 \pm 0.014$ & $344^{+52}_{-52}$ & $0.19 \pm 0.01$ \\
050730 & $3.97$ & 1.02E-07 $\pm$ 1.58E-08 & 5.80E-06 $\pm$ 2.25E-07 & $\cdots$ & $\cdots$ & $0.027 \pm 0.066$ & $124^{+26}_{-26}$ & $\cdots$ \\
050802 & $1.71$ & 5.47E-07 $\pm$ 5.32E-08 & 5.24E-06 $\pm$ 2.50E-07 & $\cdots$ & $\cdots$ & $0.070 \pm 0.036$ & $121^{+28}_{-28}$ & $2.03 \pm 1.02$ \\
050814 & $5.30$ & 1.04E-07 $\pm$ 2.24E-08 & 3.99E-06 $\pm$ 2.65E-07 & $\cdots$ & $\cdots$ & $-0.009 \pm 0.180$ & $60^{+24}_{-6}$ & $\cdots$ \\
050820A & $2.61$ & 6.12E-07 $\pm$ 3.49E-08 & 1.09E-05 $\pm$ 4.66E-07 & 6.73E-03 $\pm$ 3.09E-03 & $\cdots$ & $0.061 \pm 0.033$ & $246^{+76}_{-40}$ & $1.01 \pm 0.75$ \\
050824 & $0.83$ & 7.92E-08 $\pm$ 1.44E-08 & 7.41E-07 $\pm$ 8.81E-08 & $\cdots$ & $\cdots$ & $0.289 \pm 0.640$ & $15^{+5}_{-5}$ & $\cdots$ \\
050826 & $0.30$ & 7.66E-08 $\pm$ 1.59E-08 & 1.12E-06 $\pm$ 1.19E-07 & $\cdots$ & $\cdots$ & $0.063 \pm 0.105$ & $105^{+47}_{-47}$ & $1.11 \pm 2.28$ \\
050908 & $3.35$ & 9.83E-08 $\pm$ 1.20E-08 & 1.09E-06 $\pm$ 6.98E-08 & $\cdots$ & $\cdots$ & $-0.017 \pm 0.046$ & $41^{+9}_{-5}$ & $1.10 \pm 1.47$ \\
050922C & $2.20$ & 1.93E-06 $\pm$ 5.18E-08 & 5.09E-06 $\pm$ 1.03E-07 & $\cdots$ & $0.06 \pm 0.01$ & $0.015 \pm 0.003$ & $198^{+38}_{-22}$ & $0.13 \pm 0.01$ \\
051016B & $0.94$ & 1.92E-07 $\pm$ 1.43E-08 & 4.31E-07 $\pm$ 3.39E-08 & $\cdots$ & $\cdots$ & $0.008 \pm 0.030$ & $24^{+7}_{-7}$ & $\cdots$ \\
051022 & $0.80$ & 1.1E-5 $\pm$ 8.7E-7 & 3.40E-4 $\pm$ 1.2E-5 & 0.0029 $\pm$ 0.0001 & $\cdots$ & $0.088 \pm 0.008$ & $510^{+22}_{-20}$ & $0.19 \pm 0.04$ \\
051109A & $2.35$ & 8.30E-07 $\pm$ 8.83E-08 & 6.10E-06 $\pm$ 4.58E-07 & $\cdots$ & $\cdots$ & $-0.006 \pm 0.025$ & $161^{+130}_{-35}$ & $0.70 \pm 1.25$ \\
051111 & $1.55$ & 7.61E-07 $\pm$ 3.65E-08 & 1.38E-05 $\pm$ 2.76E-07 & $\cdots$ & $1.70 \pm 0.07$ & $0.009 \pm 0.004$ & $220^{+1703}_{-48}$ & $1.80 \pm 0.24$ \\
060108 & $2.03$ & 1.22E-07 $\pm$ 1.16E-08 & 8.62E-07 $\pm$ 5.26E-08 & $\cdots$ & $\cdots$ & $0.006 \pm 0.040$ & $65^{+600}_{-10}$ & $\cdots$ \\
060115 & $3.53$ & 1.30E-07 $\pm$ 1.09E-08 & 3.76E-06 $\pm$ 2.01E-07 & $\cdots$ & $\cdots$ & $0.019 \pm 0.029$ & $62^{+19}_{-6}$ & $1.11 \pm 1.71$ \\
060206 & $4.05$ & 4.41E-07 $\pm$ 1.63E-08 & 1.90E-06 $\pm$ 5.83E-08 & $\cdots$ & $0.01 \pm 0.03$ & $0.007 \pm 0.004$ & $78^{+23}_{-8}$ & $1.16 \pm 0.18$ \\
060210 & $3.91$ & 5.37E-07 $\pm$ 3.36E-08 & 1.97E-05 $\pm$ 6.39E-07 & $\cdots$ & $0.15 \pm 0.17$ & $0.183 \pm 0.033$ & $149^{+400}_{-35}$ & $0.73 \pm 0.50$ \\
060223A & $4.41$ & 2.06E-07 $\pm$ 1.67E-08 & 1.51E-06 $\pm$ 6.54E-08 & $\cdots$ & $\cdots$ & $0.036 \pm 0.021$ & $71^{+100}_{-10}$ & $0.41 \pm 0.23$ \\
060418 & $1.49$ & 1.49E-06 $\pm$ 4.85E-08 & 2.62E-05 $\pm$ 4.85E-07 & $\cdots$ & $0.22 \pm 0.03$ & $0.104 \pm 0.008$ & $230^{+20}_{-20}$ & $0.67 \pm 0.08$ \\
060502A & $1.51$ & 3.72E-07 $\pm$ 2.81E-08 & 6.59E-06 $\pm$ 1.77E-07 & $\cdots$ & $4.90 \pm 0.11$ & $0.004 \pm 0.010$ & $156^{+400}_{-33}$ & $2.94 \pm 1.19$ \\
060510B & $4.90$ & 9.51E-08 $\pm$ 1.12E-08 & 9.98E-06 $\pm$ 2.62E-07 & $\cdots$ & $\cdots$ & $0.110 \pm 0.060$ & $95^{+60}_{-30}$ & $\cdots$ \\
060512 & $0.44$ & 1.32E-07 $\pm$ 1.83E-08 & 6.04E-07 $\pm$ 6.34E-08 & $\cdots$ & $\cdots$ & $0.043 \pm 0.173$ & $22^{+6}_{-6}$ & $\cdots$ \\
060522 & $5.11$ & 8.73E-08 $\pm$ 1.45E-08 & 2.42E-06 $\pm$ 1.43E-07 & $\cdots$ & $\cdots$ & $0.034 \pm 0.185$ & $80^{+382}_{-12}$ & $\cdots$ \\
060526 & $3.21$ & 2.33E-07 $\pm$ 1.53E-08 & 3.01E-06 $\pm$ 2.40E-07 & 6.55E-03 $\pm$ 1.60E-03 & $0.17 \pm 0.09$ & $0.085 \pm 0.030$ & $25^{+5}_{-5}$ & $0.38 \pm 0.11$ \\
060604 & $2.68$ & 5.10E-08 $\pm$ 1.19E-08 & 9.82E-07 $\pm$ 1.57E-07 & $\cdots$ & $\cdots$ & $0.080 \pm 0.338$ & $40^{+5}_{-5}$ & $\cdots$ \\
060605 & $3.80$ & 8.56E-08 $\pm$ 1.36E-08 & 1.58E-06 $\pm$ 1.24E-07 & 8.23E-04 $\pm$ 5.14E-05 & $\cdots$ & $-0.013 \pm 0.068$ & $90^{+91}_{-12}$ & $1.22 \pm 0.72$ \\
060607A & $3.08$ & 2.66E-07 $\pm$ 1.50E-08 & 6.33E-06 $\pm$ 1.69E-07 & $\cdots$ & $1.98 \pm 0.11$ & $0.025 \pm 0.008$ & $120^{+190}_{-17}$ & $1.23 \pm 0.68$ \\
060707 & $3.43$ & 1.53E-07 $\pm$ 2.12E-08 & 3.41E-06 $\pm$ 1.96E-07 & $\cdots$ & $\cdots$ & $0.050 \pm 0.054$ & $63^{+13}_{-6}$ & $\cdots$ \\
060714 & $2.71$ & 2.30E-07 $\pm$ 1.42E-08 & 6.88E-06 $\pm$ 2.47E-07 & $\cdots$ & $\cdots$ & $0.125 \pm 0.022$ & $103^{+21}_{-16}$ & $\cdots$ \\
060729 & $0.54$ & 1.93E-07 $\pm$ 1.30E-08 & 6.43E-06 $\pm$ 3.16E-07 & $\cdots$ & $\cdots$ & $0.092 \pm 0.041$ & $61^{+9}_{-9}$ & $\cdots$ \\
060814 & $0.84$ & 1.83E-06 $\pm$ 4.44E-08 & 4.94E-05 $\pm$ 4.91E-07 & $\cdots$ & $0.29 \pm 0.03$ & $0.040 \pm 0.003$ & $257^{+74}_{-35}$ & $1.65 \pm 0.24$ \\
060904B & $0.70$ & 4.37E-07 $\pm$ 2.28E-08 & 4.05E-06 $\pm$ 2.17E-07 & $\cdots$ & $0.36 \pm 0.09$ & $0.003 \pm 0.008$ & $80^{+770}_{-12}$ & $1.00 \pm 0.16$ \\
060908 & $2.43$ & 6.69E-07 $\pm$ 3.36E-08 & 7.68E-06 $\pm$ 1.85E-07 & $\cdots$ & $0.26 \pm 0.06$ & $0.061 \pm 0.008$ & $151^{+112}_{-25}$ & $0.52 \pm 0.09$ \\
060926 & $3.21$ & 1.56E-07 $\pm$ 1.22E-08 & 5.47E-07 $\pm$ 3.80E-08 & $\cdots$ & $1.03 \pm 0.11$ & $0.148 \pm 0.050$ & $20^{+11}_{-11}$ & $\cdots$ \\
060927 & $5.60$ & 4.02E-07 $\pm$ 1.54E-08 & 2.37E-06 $\pm$ 8.67E-08 & $\cdots$ & $0.12 \pm 0.04$ & $0.094 \pm 0.010$ & $72^{+15}_{-7}$ & $0.46 \pm 0.12$ \\
061007 & $1.26$ & 7.20E-06 $\pm$ 1.11E-07 & 2.24E-04 $\pm$ 1.72E-06 & $\cdots$ & $0.11 \pm 0.01$ & $0.066 \pm 0.003$ & $399^{+12}_{-11}$ & $0.38 \pm 0.02$ \\
061110A & $0.76$ & 9.79E-08 $\pm$ 1.35E-08 & 2.71E-06 $\pm$ 1.18E-07 & $\cdots$ & $\cdots$ & $-0.038 \pm 0.050$ & $90^{+13}_{-13}$ & $\cdots$ \\
061110B & $3.44$ & 1.79E-07 $\pm$ 2.66E-08 & 6.12E-06 $\pm$ 3.38E-07 & $\cdots$ & $0.24 \pm 0.36$ & $0.155 \pm 0.064$ & $517^{+53}_{-53}$ & $0.79 \pm 0.64$ \\
061121 & $1.31$ & 8.04E-06 $\pm$ 1.07E-07 & 6.53E-05 $\pm$ 5.76E-07 & $\cdots$ & $0.03 \pm 0.01$ & $0.050 \pm 0.003$ & $606^{+55}_{-44}$ & $0.98 \pm 0.19$ \\
061222B & $3.36$ & 2.29E-07 $\pm$ 3.15E-08 & 5.01E-06 $\pm$ 2.49E-07 & $\cdots$ & $\cdots$ & $0.024 \pm 0.043$ & $49^{+8}_{-8}$ & $\cdots$ \\
070110 & $2.35$ & 1.12E-07 $\pm$ 1.36E-08 & 4.04E-06 $\pm$ 1.64E-07 & $\cdots$ & $\cdots$ & $-0.010 \pm 0.031$ & $110^{+30}_{-30}$ & $\cdots$ \\
070208 & $1.17$ & 1.39E-07 $\pm$ 2.06E-08 & 1.06E-06 $\pm$ 1.46E-07 & $\cdots$ & $\cdots$ & $0.083 \pm 0.211$ & $51^{+10}_{-10}$ & $\cdots$ \\
070318 & $0.84$ & 4.10E-07 $\pm$ 2.12E-08 & 7.34E-06 $\pm$ 2.01E-07 & $\cdots$ & $\cdots$ & $0.037 \pm 0.008$ & $154^{+19}_{-19}$ & $0.72 \pm 0.24$ \\
070411 & $2.95$ & 1.50E-07 $\pm$ 1.31E-08 & 6.29E-06 $\pm$ 2.19E-07 & $\cdots$ & $\cdots$ & $0.041 \pm 0.029$ & $83^{+11}_{-11}$ & $\cdots$ \\
070506 & $2.31$ & 1.67E-07 $\pm$ 1.38E-08 & 5.16E-07 $\pm$ 3.43E-08 & $\cdots$ & $2.52 \pm 0.04$ & $0.010 \pm 0.030$ & $31^{+2}_{-3}$ & $0.12 \pm 0.06$ \\
070508 & $0.82$ & 7.67E-06 $\pm$ 1.18E-07 & 7.26E-05 $\pm$ 6.15E-07 & $\cdots$ & $0.04 \pm 0.01$ & $0.106 \pm 0.003$ & $233^{+7}_{-7}$ & $0.20 \pm 0.01$ \\
070521 & $0.55$ & 2.09E-06 $\pm$ 5.26E-08 & 2.97E-05 $\pm$ 4.00E-07 & $\cdots$ & $0.04 \pm 0.01$ & $0.116 \pm 0.004$ & $222^{+16}_{-12}$ & $0.58 \pm 0.06$ \\
070529 & $2.50$ & 3.32E-07 $\pm$ 5.08E-08 & 7.44E-06 $\pm$ 4.31E-07 & $\cdots$ & $\cdots$ & $0.170 \pm 0.091$ & $180^{+52}_{-52}$ & $\cdots$ \\
070611 & $2.04$ & 1.45E-07 $\pm$ 2.25E-08 & 9.52E-07 $\pm$ 8.44E-08 & $\cdots$ & $\cdots$ & $0.053 \pm 0.080$ & $92^{+30}_{-30}$ & $\cdots$ \\
070612A & $0.62$ & 2.77E-07 $\pm$ 4.24E-08 & 2.72E-05 $\pm$ 9.37E-07 & $\cdots$ & $\cdots$ & $0.032 \pm 0.023$ & $87^{+17}_{-17}$ & $2.49 \pm 1.48$ \\
070714B & $0.92$ & 3.24E-06 $\pm$ 1.46E-07 & 8.91E-06 $\pm$ 6.77E-07 & $\cdots$ & $0.03 \pm 0.01$ & $0.164 \pm 0.021$ & $1120^{+473}_{-230}$ & $0.45 \pm 0.04$ \\
070802 & $2.45$ & 6.38E-08 $\pm$ 9.69E-09 & 6.50E-07 $\pm$ 7.05E-08 & $\cdots$ & $\cdots$ & $-0.156 \pm 0.150$ & $70^{+25}_{-25}$ & $\cdots$ \\
070810A & $2.17$ & 2.77E-07 $\pm$ 1.77E-08 & 1.59E-06 $\pm$ 8.43E-08 & $\cdots$ & $1.09 \pm 0.23$ & $-0.006 \pm 0.015$ & $44^{+9}_{-9}$ & $0.73 \pm 0.22$ \\
071003 & $1.10$ & 4.71E-06 $\pm$ 1.82E-07 & 6.73E-05 $\pm$ 1.48E-06 & $\cdots$ & $0.38 \pm 0.05$ & $0.072 \pm 0.007$ & $799^{+75}_{-61}$ & $0.88 \pm 0.07$ \\
071010A & $0.98$ & 1.17E-07 $\pm$ 2.67E-08 & 4.97E-07 $\pm$ 6.05E-08 & $\cdots$ & $\cdots$ & $-0.076 \pm 0.153$ & $27^{+10}_{-10}$ & $\cdots$ \\
071010B & $0.95$ & 9.20E-07 $\pm$ 2.18E-08 & 8.37E-06 $\pm$ 1.16E-07 & $\cdots$ & $0.84 \pm 0.04$ & $0.010 \pm 0.003$ & $52^{+6}_{-8}$ & $1.21 \pm 0.03$ \\
071031 & $2.69$ & 7.08E-08 $\pm$ 8.61E-09 & 2.19E-06 $\pm$ 1.92E-07 & $\cdots$ & $\cdots$ & $-0.038 \pm 0.108$ & $24^{+7}_{-7}$ & $\cdots$ \\
071117 & $1.33$ & 2.71E-06 $\pm$ 5.83E-08 & 7.97E-06 $\pm$ 2.02E-07 & $\cdots$ & $0.60 \pm 0.01$ & $0.009 \pm 0.003$ & $278^{+143}_{-48}$ & $0.20 \pm 0.02$ \\
071122 & $1.14$ & 6.76E-08 $\pm$ 2.06E-08 & 1.41E-06 $\pm$ 1.63E-07 & $\cdots$ & $\cdots$ & $0.391 \pm 0.392$ & $73^{+30}_{-30}$ & $\cdots$ \\
080210 & $2.64$ & 2.57E-07 $\pm$ 1.95E-08 & 4.17E-06 $\pm$ 1.41E-07 & $\cdots$ & $0.53 \pm 0.17$ & $0.019 \pm 0.013$ & $73^{+15}_{-15}$ & $0.57 \pm 0.44$ \\
080310 & $2.43$ & 1.83E-07 $\pm$ 1.72E-08 & 5.49E-06 $\pm$ 2.90E-07 & $\cdots$ & $\cdots$ & $0.038 \pm 0.021$ & $28^{+6}_{-6}$ & $0.41 \pm 0.55$ \\
080319B & $0.94$ & 1.55E-05 $\pm$ 1.91E-07 & 5.25E-04 $\pm$ 3.94E-06 & $\cdots$ & $0.02 \pm 0.01$ & $0.031 \pm 0.003$ & $651^{+8}_{-8}$ & $0.14 \pm 0.01$ \\
080319C & $1.95$ & 2.22E-06 $\pm$ 7.79E-08 & 1.77E-05 $\pm$ 2.99E-07 & $\cdots$ & $\cdots$ & $0.042 \pm 0.007$ & $307^{+85}_{-56}$ & $0.21 \pm 0.12$ \\
080330 & $1.51$ & 1.33E-07 $\pm$ 1.80E-08 & 8.77E-07 $\pm$ 1.26E-07 & $\cdots$ & $\cdots$ & $0.109 \pm 0.060$ & $20^{+9}_{-9}$ & $\cdots$ \\
080411 & $1.03$ & 1.04E-05 $\pm$ 1.31E-07 & 8.75E-05 $\pm$ 2.01E-07 & $\cdots$ & $0.21 \pm 0.01$ & $0.167 \pm 0.003$ & $259^{+21}_{-16}$ & $0.65 \pm 0.01$ \\
080413A & $2.43$ & 1.22E-06 $\pm$ 2.65E-08 & 9.86E-06 $\pm$ 1.71E-07 & $\cdots$ & $0.13 \pm 0.03$ & $0.078 \pm 0.004$ & $170^{+48}_{-24}$ & $0.23 \pm 0.03$ \\
080413B & $1.10$ & 3.17E-06 $\pm$ 8.25E-08 & 8.00E-06 $\pm$ 1.52E-07 & $\cdots$ & $0.23 \pm 0.01$ & $0.004 \pm 0.003$ & $73^{+10}_{-10}$ & $0.50 \pm 0.03$ \\
080430 & $0.77$ & 4.60E-07 $\pm$ 2.15E-08 & 3.01E-06 $\pm$ 1.53E-07 & $\cdots$ & $0.68 \pm 0.08$ & $0.009 \pm 0.004$ & $80^{+15}_{-15}$ & $0.76 \pm 0.12$ \\
080516 & $3.20$ & 2.77E-07 $\pm$ 2.80E-08 & 5.88E-07 $\pm$ 5.50E-08 & $\cdots$ & $0.15 \pm 0.01$ & $0.168 \pm 0.055$ & $66^{+24}_{-24}$ & $\cdots$ \\
080520 & $1.55$ & 8.23E-08 $\pm$ 1.00E-08 & 1.59E-07 $\pm$ 3.00E-08 & $\cdots$ & $\cdots$ & $0.037 \pm 0.098$ & $12^{+5}_{-5}$ & $\cdots$ \\
080603B & $2.69$ & 7.57E-07 $\pm$ 2.63E-08 & 7.02E-06 $\pm$ 1.78E-07 & $\cdots$ & $0.08 \pm 0.01$ & $0.283 \pm 0.010$ & $85^{+55}_{-18}$ & $0.22 \pm 0.03$ \\
080605 & $1.64$ & 5.99E-06 $\pm$ 1.10E-07 & 4.72E-05 $\pm$ 4.32E-07 & $\cdots$ & $0.11 \pm 0.01$ & $0.057 \pm 0.003$ & $246^{+14}_{-11}$ & $0.22 \pm 0.01$ \\
080607 & $3.04$ & 8.35E-06 $\pm$ 2.42E-07 & 1.00E-04 $\pm$ 0.00E+00 & $\cdots$ & $0.04 \pm 0.01$ & $0.035 \pm 0.003$ & $394^{+35}_{-33}$ & $0.18 \pm 0.06$ \\
080707 & $1.23$ & 1.68E-07 $\pm$ 1.02E-08 & 1.26E-06 $\pm$ 8.87E-08 & $\cdots$ & $\cdots$ & $0.093 \pm 0.032$ & $73^{+20}_{-20}$ & $\cdots$ \\
080721 & $2.60$ & 9.57E-06 $\pm$ 5.01E-07 & 5.99E-05 $\pm$ 3.04E-06 & $\cdots$ & $0.13 \pm 0.05$ & $0.048 \pm 0.009$ & $485^{+41}_{-36}$ & $0.09 \pm 0.04$ \\
090423 & $8.2$ & 2.17E-07 $\pm$ 1.55E-08 & 1.15E-06 $\pm$ 4.73E-08 & $\cdots$ & $\cdots$ & $\cdots$ & $48.6^{+3.8}_{-3.8}$ & $\cdots$ \\
  \hline
  \caption{The data of 116 GRBs used in our analysis.
    For pre-\emph{Swift} GRBs, we take the values of $P_{\rm bolo}$ and
    $S_{\rm bolo}$ directly from Schaefer (2007). For those GRBs observed
    by \emph{Swift}, we adopt the values of $P$ and $S$ from \emph{Swift}
    website and calculate $P_{\rm bolo}$ and $S_{\rm bolo}$.
    We use the $F_{\rm beam}$ value from Ghirlanda et al. (2007). Other
    data are taken from Xiao \& Schaefer (2009).}
  \label{tab:grbresults}
\end{longtable}
}
\twocolumn

The six luminosity relations can be expressed, in general,
as $R=A Q^{b}$ and
Eq.~(\ref{eq:GRB-lag-L})-(\ref{eq:GRB-E_peak-E_iso}) are the
corresponding logarithm forms
\be
\log{R} = \log{A} + b \log{Q} \Rightarrow  y= a + b x.
\label{rab}
\ee
For the fit of this linear relation, we used the techniques presented
in D'Agostini (2005), according to which, the joint likelihood
function for the coefficients $a$ and $b$ and the intrinsic scatter
$\sigma_{\mathrm{int}}$ is
\begin{eqnarray}
\label{eq:likelihood} L(a,b,\sigma _{{\mathop{\rm int}} } ) \propto
\prod\limits_i {\frac{1}{{\sqrt {\sigma ^2 _{{\mathop{\rm int}} }  +
\sigma ^2 _{y_i }  + b^2 \sigma ^2 _{x_i } } }}}\nonumber \\  \times
\exp [ - \frac{{(y_i - a - bx_i )^2 }}{{2(\sigma ^2 _{{\mathop{\rm
int}} }  + \sigma ^2 _{y_i }  + b^2 \sigma ^2 _{x_i } )}}]
\end{eqnarray}
where $x_i$ and $y_i$ are corresponding observational data for the
$i$th GRB.
When considering error propagation from a quantity, say $\xi$ with
error $\sigma_{\xi}$, to its logarithm, we set $ \frac {
  \log(\xi + \sigma_{\xi}^{+})
  +
  \log(\xi - \sigma_{\xi}^{-})
} {
  2
} $ and $ \frac {
  \log(\xi + \sigma_{\xi}^{+})
  -
  \log(\xi - \sigma_{\xi}^{-})
} {
  2
} $ as the center value and the error of the logarithm
correspondingly. This requires $\xi > \sigma_{\xi}^{-}$ (the
quantities we are interested in here are all positive). Due to the
limitation of the data, for a given luminosity correlation, not all
the GRBs have all of the needed observational quantities available
and satisfy $\xi > \sigma_{\xi}^{-}$ at the same time. The numbers
of GRBs for each fit of the luminosity correlations are included in
Table~\ref{tablebs2007}.

\section{Test of the updated luminosity correlations}

\subsection{Luminosity correlations}

\begin{figure*}
  \includegraphics[width = 0.45 \textwidth]{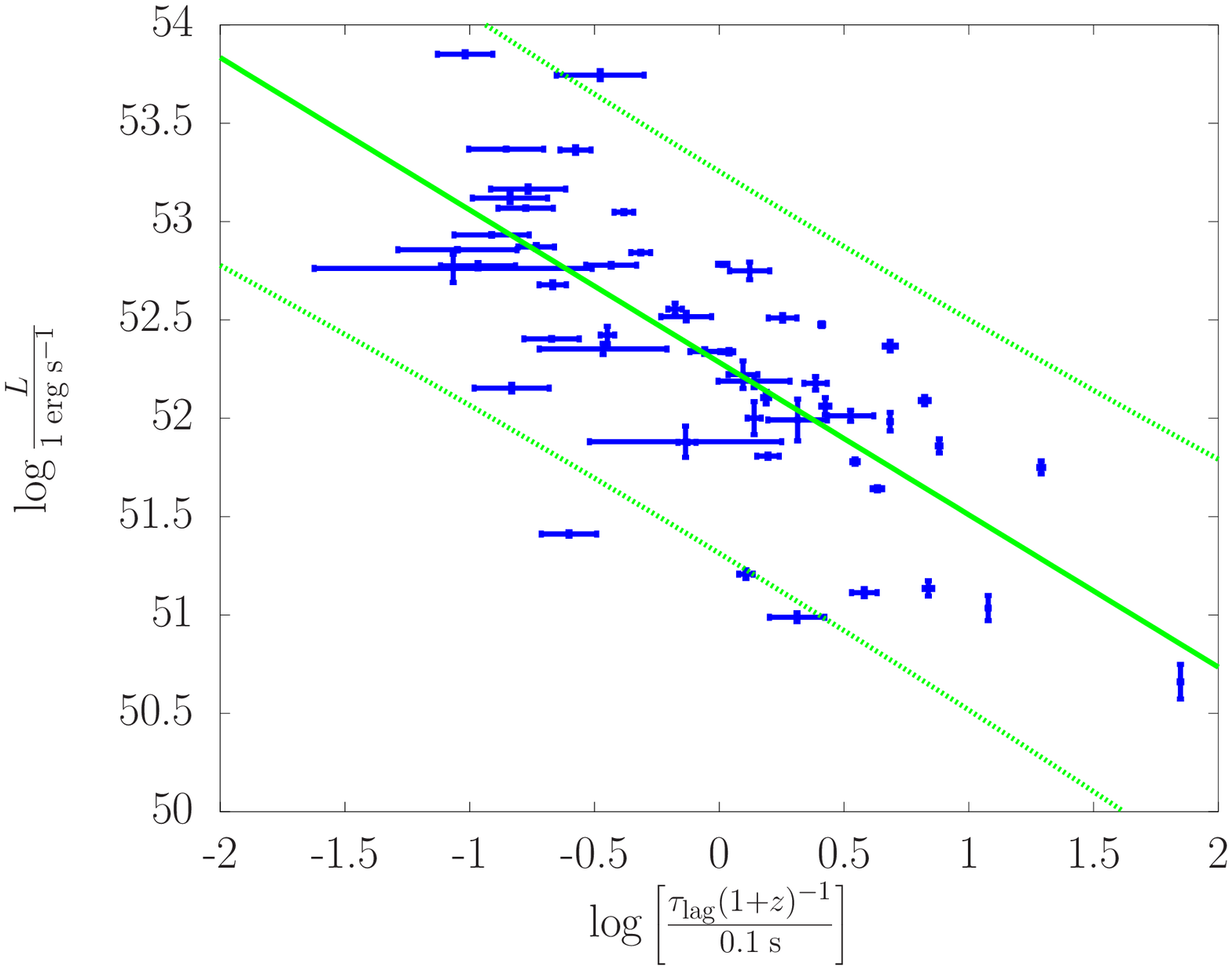}
  \includegraphics[width = 0.45 \textwidth]{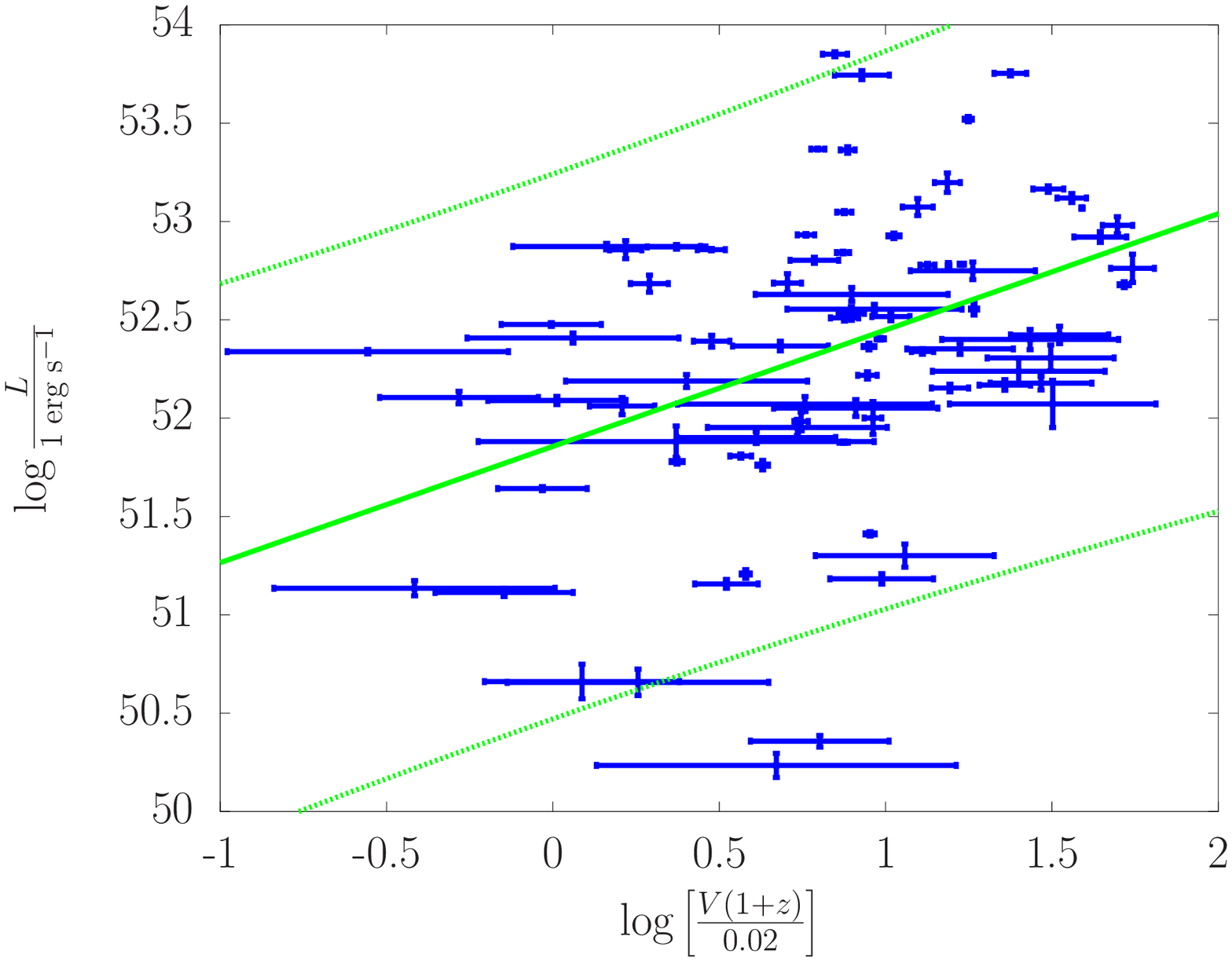}
  \\
  \includegraphics[width = 0.45 \textwidth]{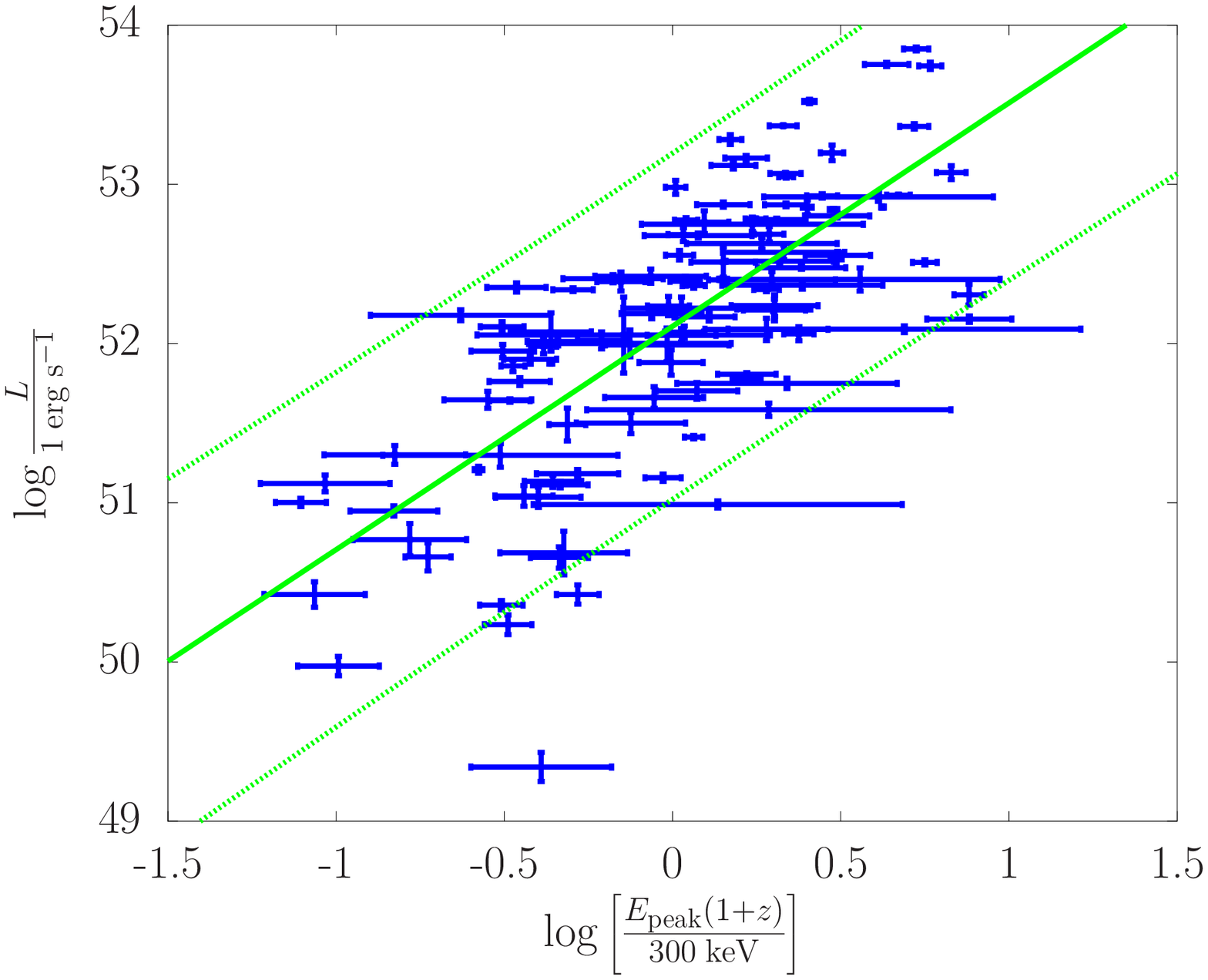}
  \includegraphics[width = 0.45 \textwidth]{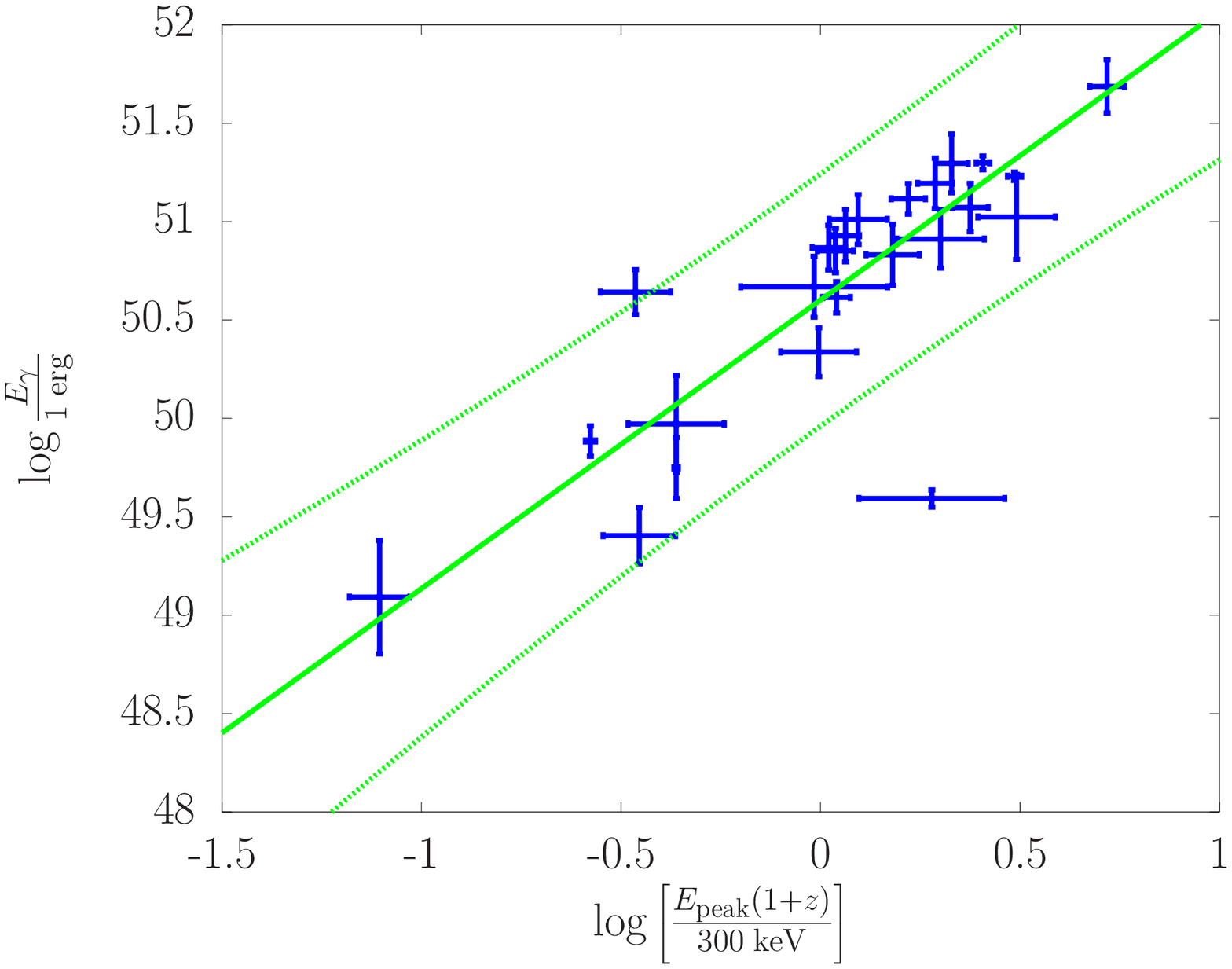}
  \\
  \includegraphics[width = 0.45 \textwidth]{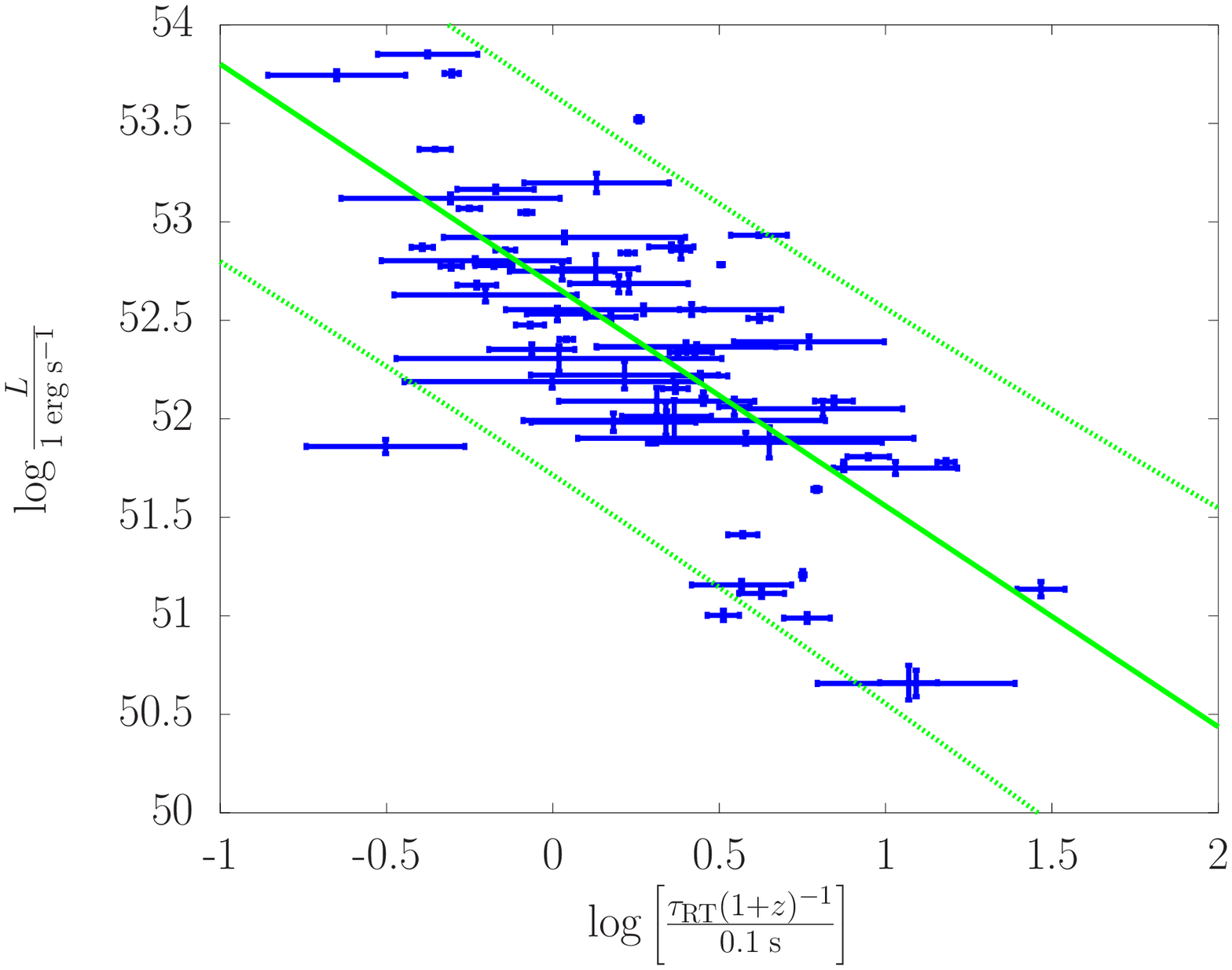}
  \includegraphics[width = 0.45 \textwidth]{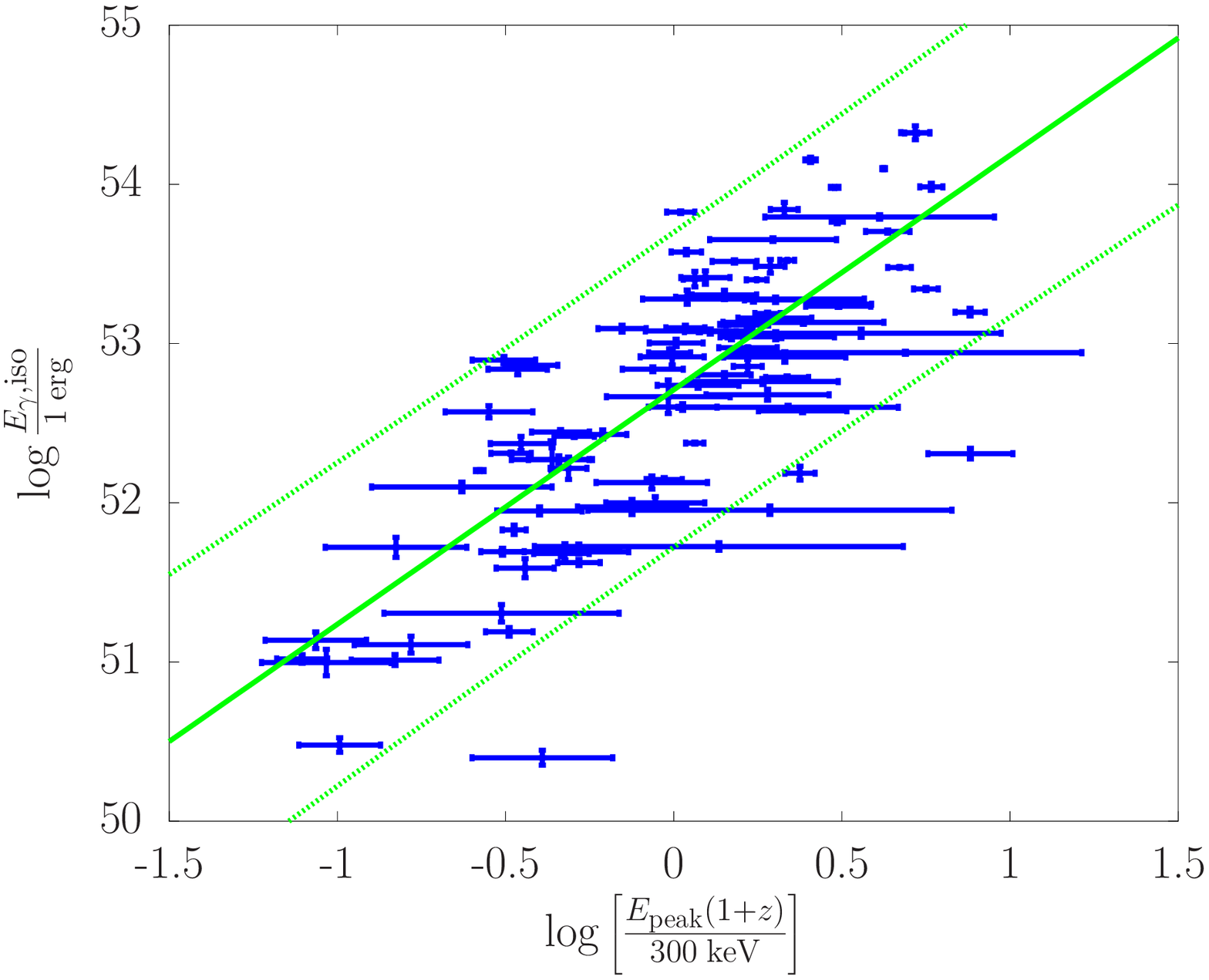}
  \caption{\label{Fig1} The $\tlag-L$, $V-L$, $\epkk-L$,
    $\epkk-E_\gamma$, $\trt-L$ and $\epkk-E_{\gamma, \mathrm{iso}}$
    correlations.
    The $1\sigma$ uncertainties are used as the error bar. The solid
    lines show the best fit results. The dotted lines plot the
    $2 \sigma$ confidence regions.}
  \label{fig:fit}
\end{figure*}

Our fitting results for the six luminosity correlations are shown in
Figure~\ref{fig:fit} and the last column of Table~\ref{tablebs2007}.
We assume a flat $\Lambda$CDM
with $\omm =0.27$ and $H_0=70$~km~s$^{-1}$Mpc$^{-1}$ obtained from
the five years WMAP data (Komatsu et al. 2009). The best-fit line
and $2\sigma$ confidence region are plotted in Figure~\ref{fig:fit}.

From Figure~\ref{fig:fit}, we can see that the $\epkk-E_\gamma$
correlation is the tightest one. The $V-L$ relation is quite
scattered. Its intrinsic scatter ($\sigma_{\rm int}=0.67$) has been
larger than the one that could be expected for a linear relation.

\subsection{Test redshift variation of correlations}

In order to test if the correlations discussed in the above section
vary with redshift, we divide the GRB samples into four groups
corresponding to the following redshift bins: $z\in [0, 1]$,
$z\in (1, 2]$, $z\in (2, 3]$ and $z\in (3,8.5]$.
For each correlation and each redshift bin, we perform the same fit
procedure as applied to the whole GRB sample to determine the
parameters $a$, $b$ and the intrinsic scatter $\sigma_{\rm int}$.
The results of the fits and the number of GRBs used in each fit are
summarized in Table~\ref{tablebs2007}.

\begin{figure*}
  \includegraphics[width = 0.3 \textwidth]{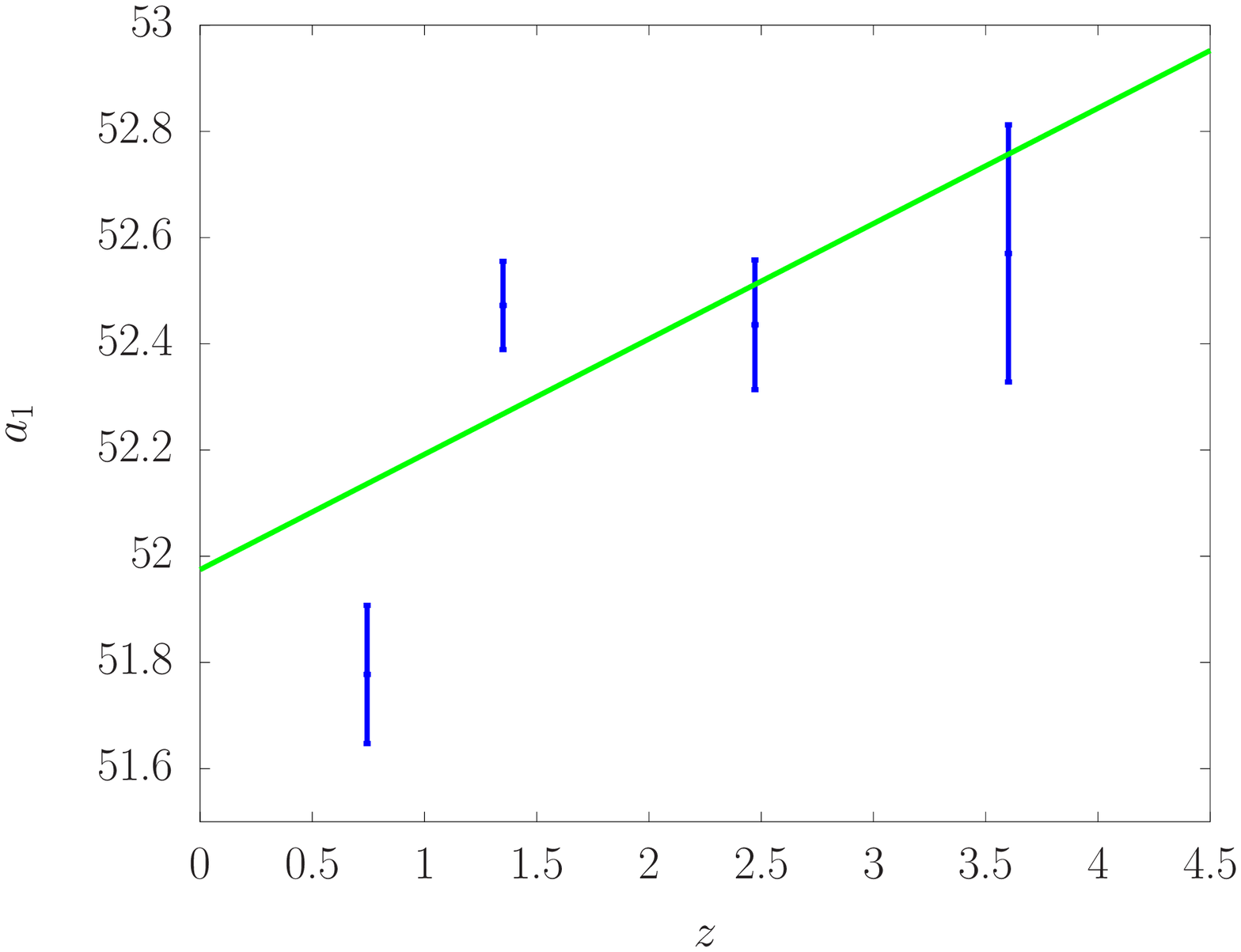}
  \includegraphics[width = 0.3 \textwidth]{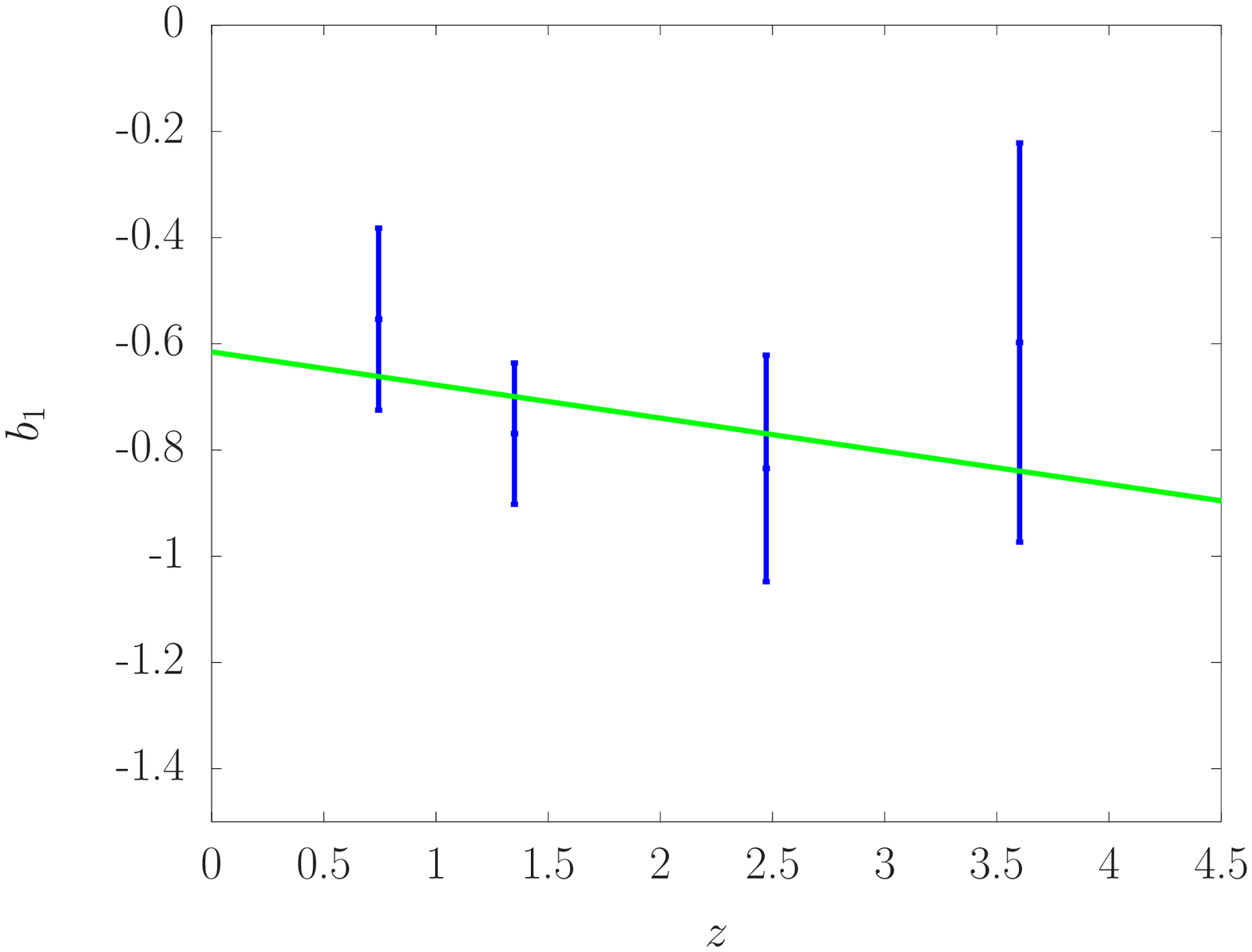}
  \\
  \includegraphics[width = 0.3 \textwidth]{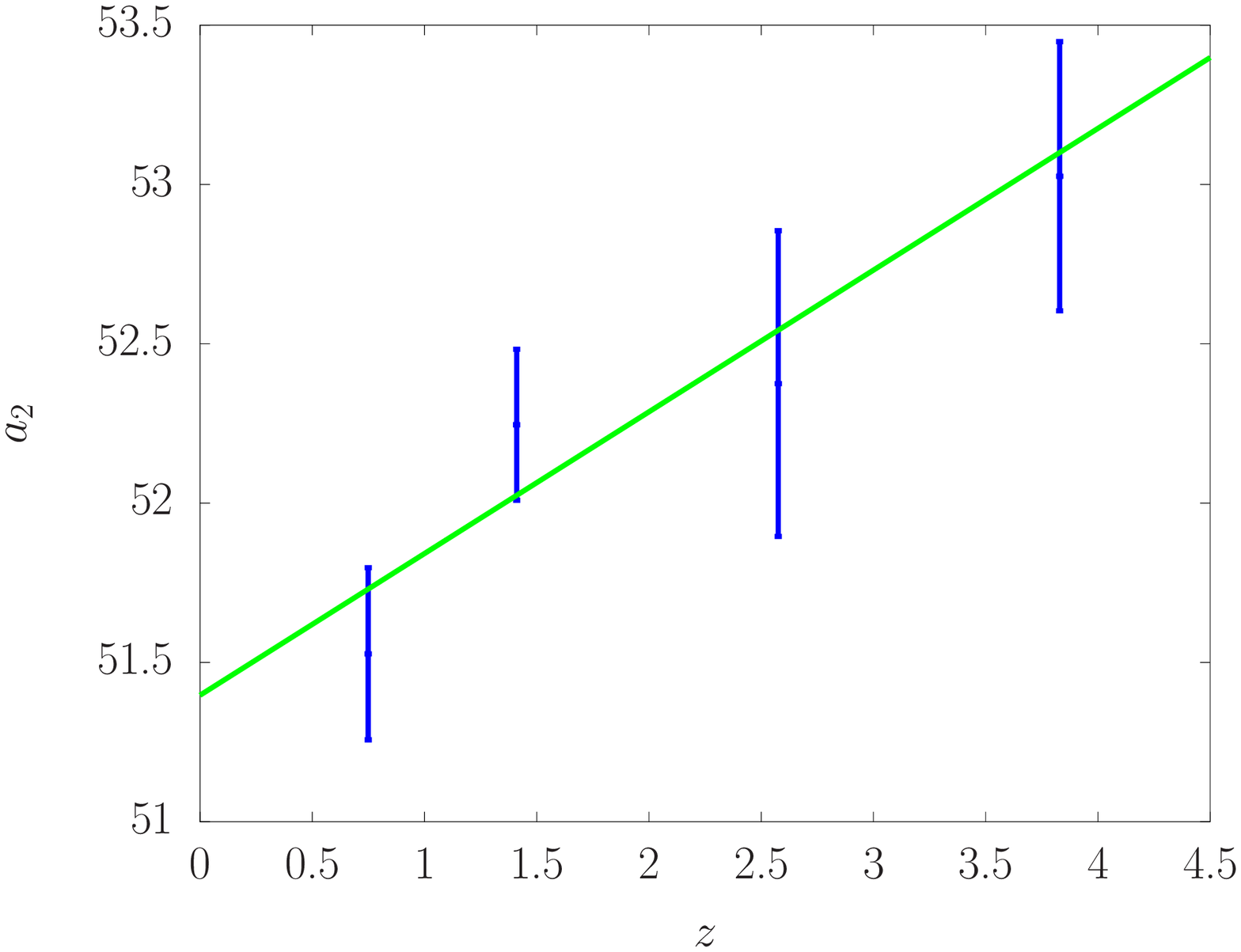}
  \includegraphics[width = 0.3 \textwidth]{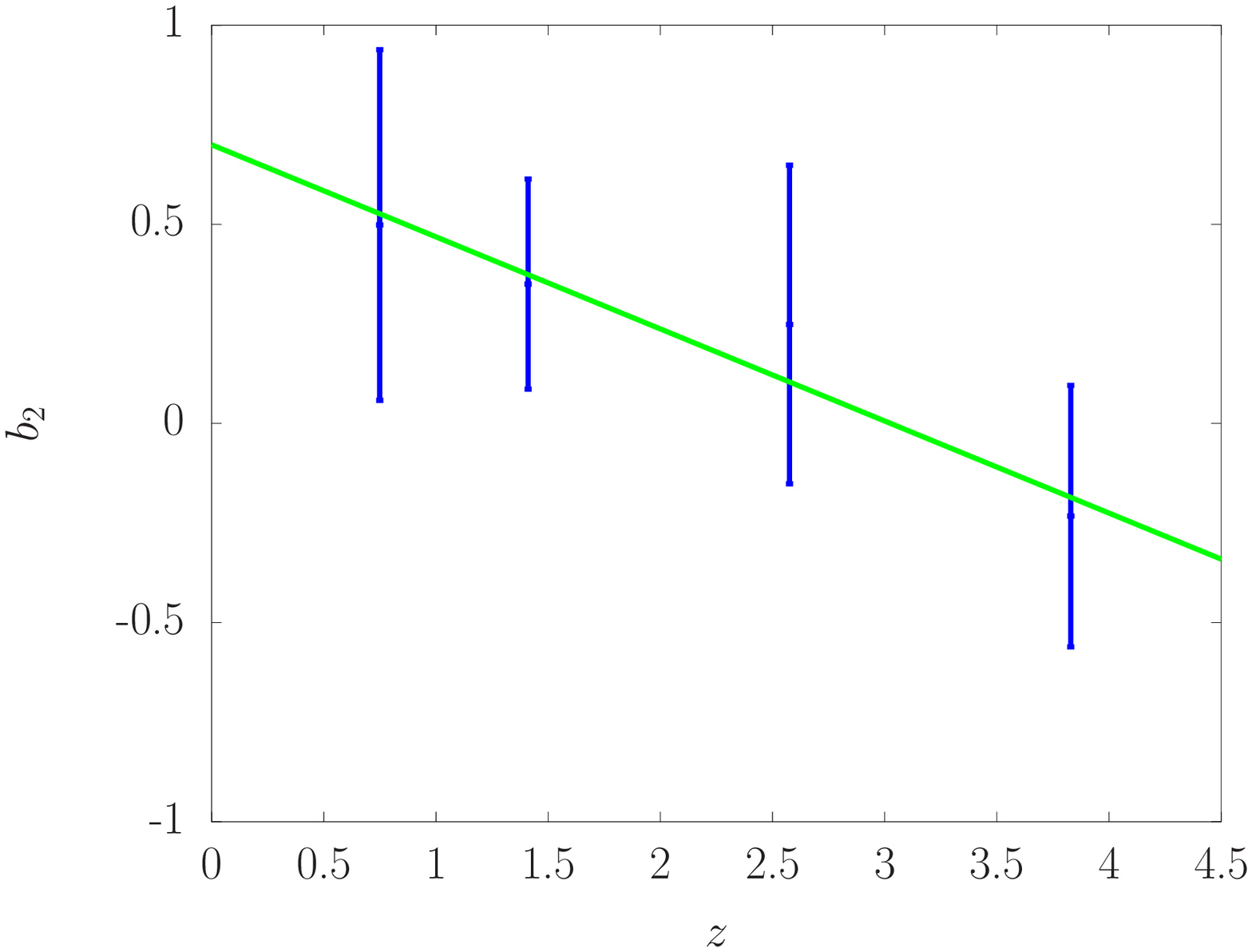}
  \\
  \includegraphics[width = 0.3 \textwidth]{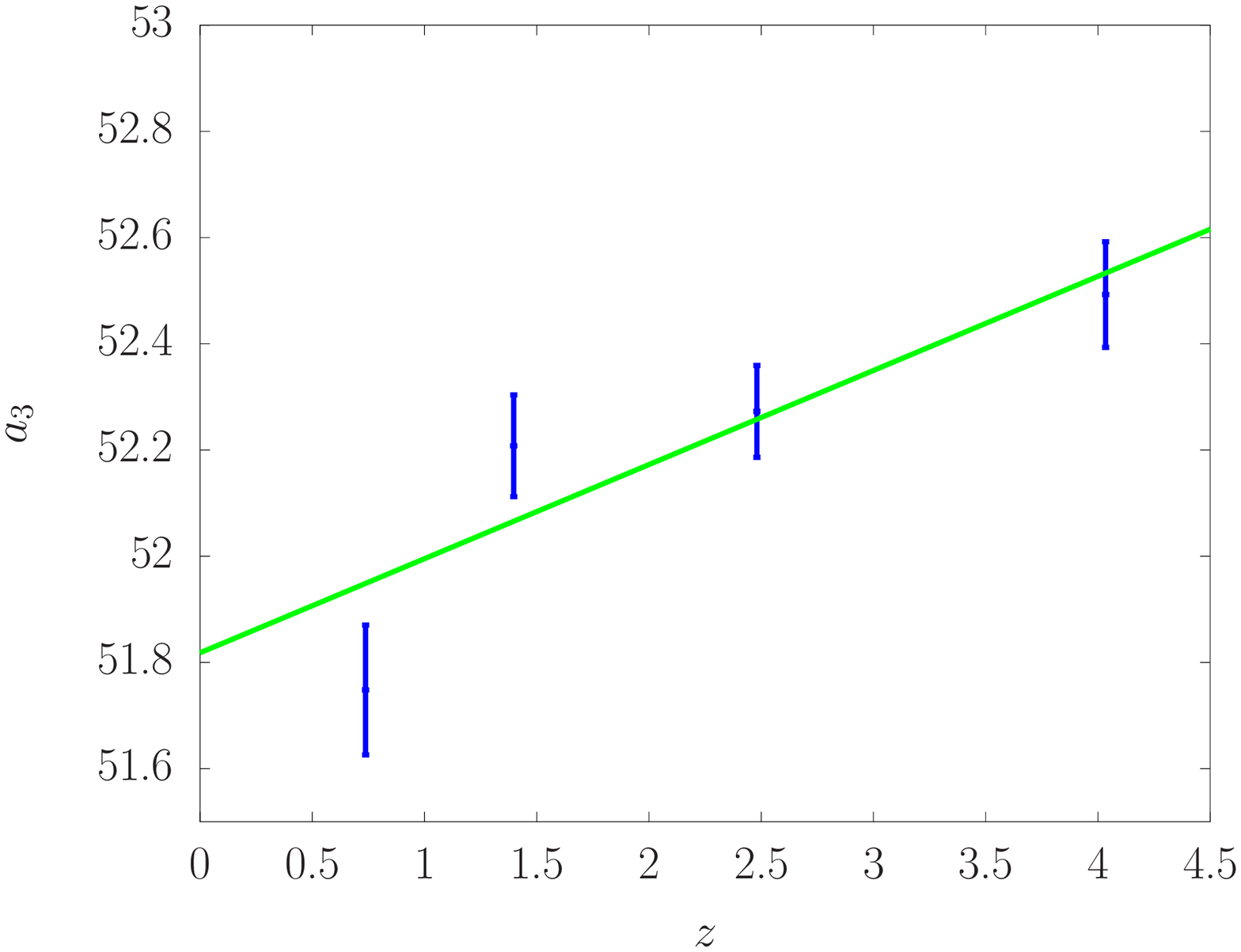}
  \includegraphics[width = 0.3 \textwidth]{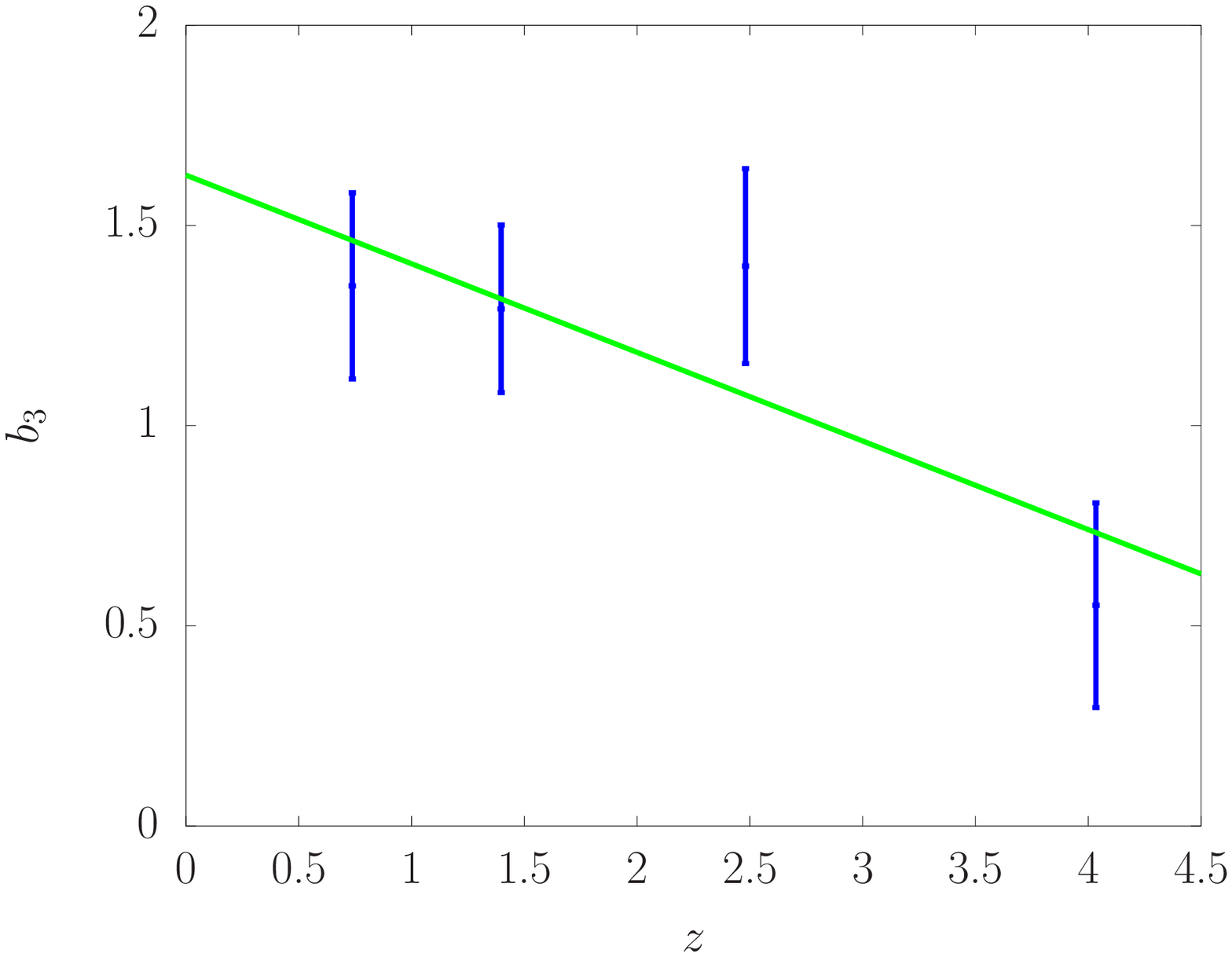}
  \\
  \includegraphics[width = 0.3 \textwidth]{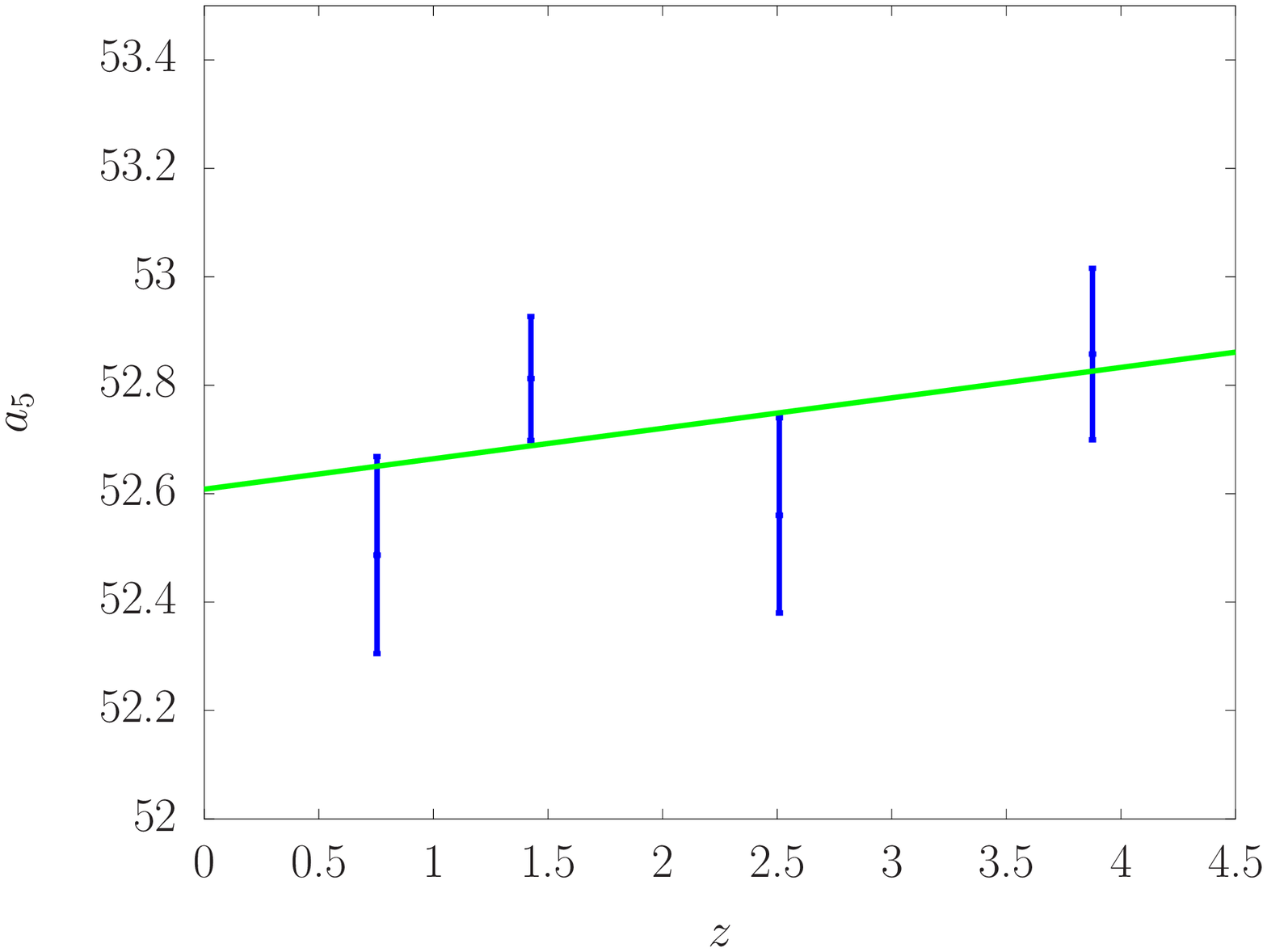}
  \includegraphics[width = 0.3 \textwidth]{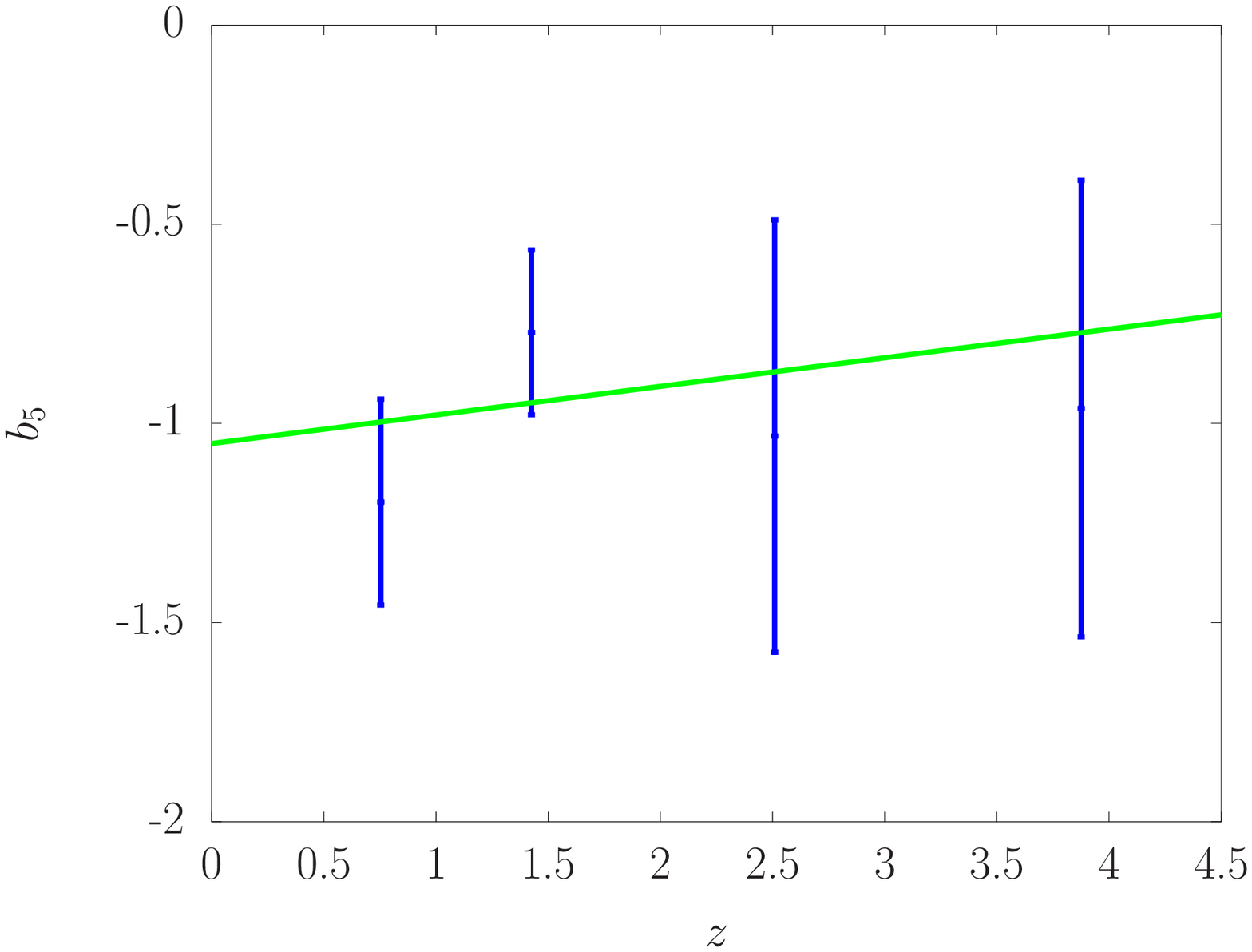}
  \\
  \includegraphics[width = 0.3 \textwidth]{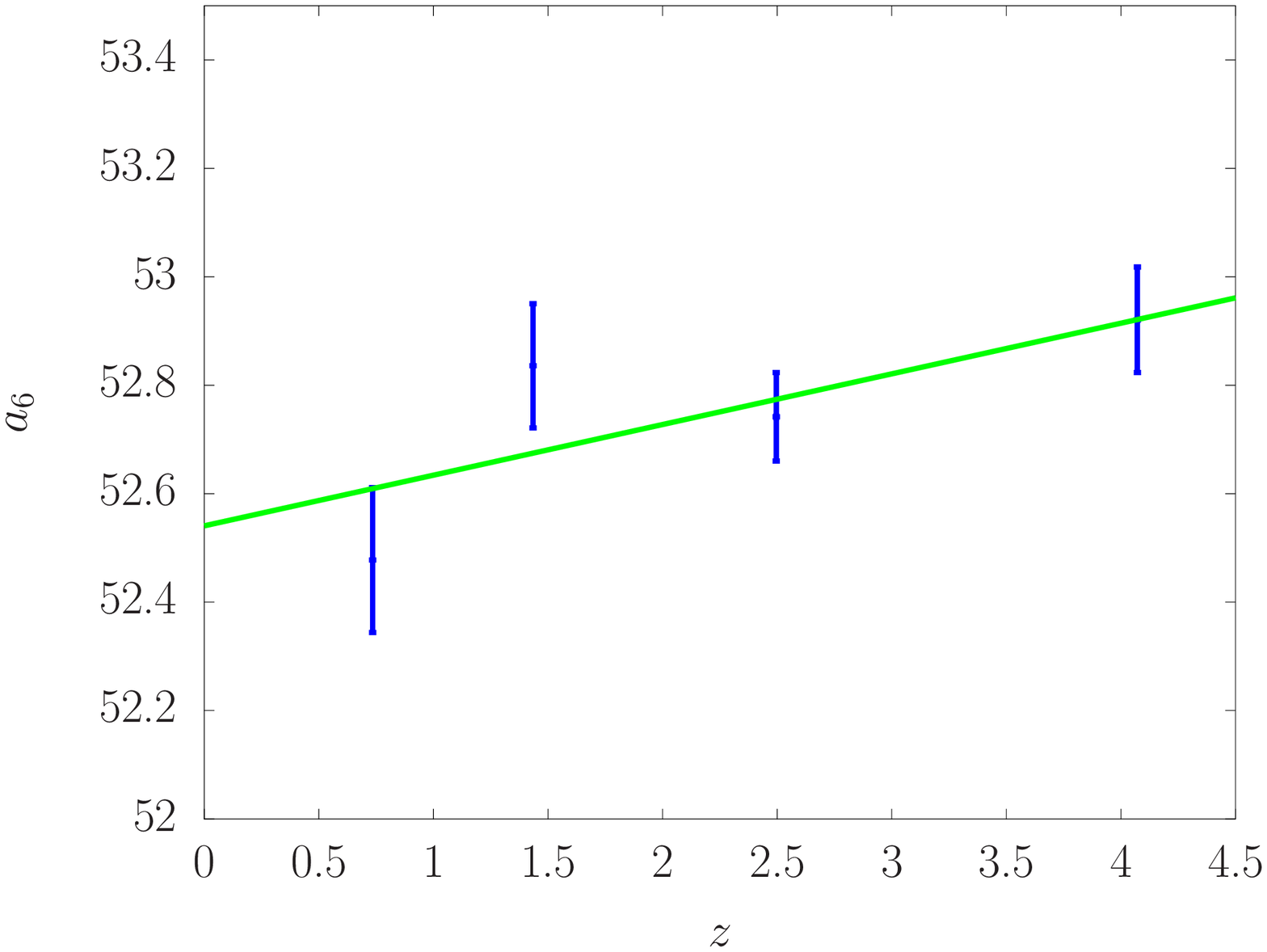}
  \includegraphics[width = 0.3 \textwidth]{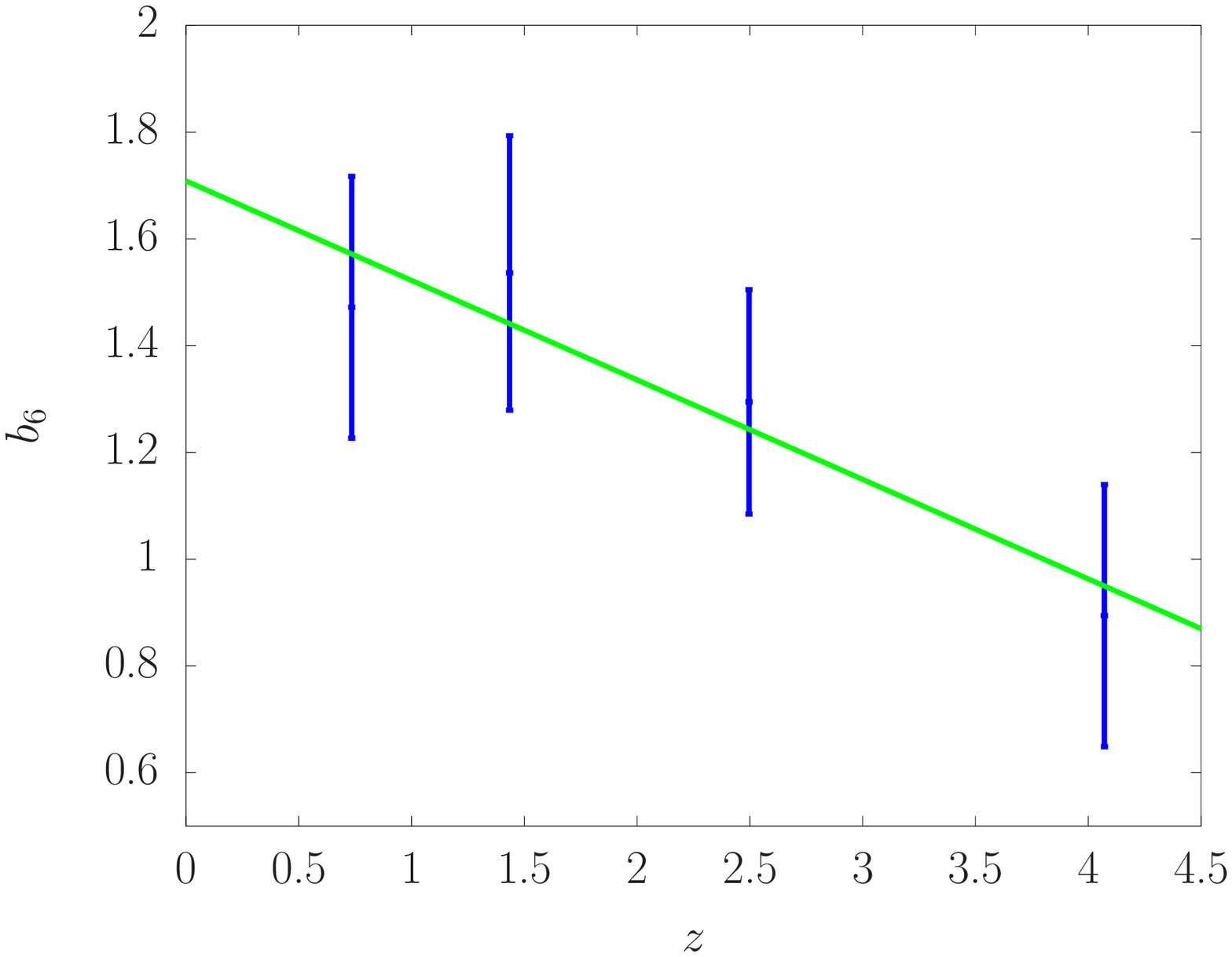}
  \caption{\label{Fig2} The correlation coefficients $a,b$ obtained in
    four redshift bins for the five correlations. The $\epkk-E_\gamma$
    correlation are not included because there are no enough GRBs in
    each redshift bin for this correlation.}
\end{figure*}

For further analysis, we perform linear fits to the parameters $a$
and $b$ versus redshift (the redshifts for the parameters $a$ and
$b$ are calculated just by averaging the redshifts of the GRBs used
in deriving corresponding $a$ and $b$). These fits are shown in
Figure~\ref{Fig2} and the slopes of $a$ and $b$ versus redshift are
presented in Table~\ref{tab:ab_slope}. For the
$E_{\mathrm{peak}}-E_{\gamma}$ correlation, there are no enough GRB
samples to perform such fits. For the other luminosity correlations
except for $E_{\mathrm{peak}}-L$, the slopes of $b$ versus redshift
are all consistent with zero at the $2\sigma$ confidence level, and
even for $E_{\mathrm{peak}}-L$ correlation, zero is near the edge of
the $2\sigma$ confidence interval of the slope of $b$ versus redshift.
Considering that, for the redshifts corresponding to $a$ and $b$, we
only loosely use the average values of the redshifts of
corresponding GRBs and the uncertainties in the redshifts are not
taken into account (which leads to an underestimate of the
uncertainties in the slopes), we can conclude that there is no
statistically significant evidence for the evolution of the
luminosity correlations with redshift. We didn't take into account
the redshift evolution of the parameters $a$ when drawing the
conclusion, since they correspond to normalization factors in the
luminosity correlations and a small change in $b$ may lead to a
larger change in $a$. In fact, as can be seen from
Table~\ref{tab:ab_slope}, the slopes of $a$ versus redshift for the
first three luminosity correlations considerably deviate from zero.

\begin{table*}
 \begin{tabular}{c c c c c c c}
\hline
   Correlation & $z\in [0,1]$  & $z\in [1,2]$ & $z\in [2,3]$ & $z\in [3,8.5]$ & Total \\
\hline
\\
$\tlag-L$&$a=51.78\pm 0.13$&$a=52.47\pm 0.08$ & $a=52.44\pm 0.12$& $a=52.57\pm 0.24$& $a=52.28\pm 0.07$ \\
  &$b=-0.55\pm 0.17$ &$b=-0.77\pm0.13$ & $b=-0.83\pm 0.21$ & $b=-0.60\pm 0.38$  & $b=-0.77\pm 0.10$  \\
  $\sigma_{\rm int}$ & $0.47\pm 0.11$ & $0.34\pm 0.07$ & $0.36\pm 0.12$ & $0.56\pm 0.26$ & $0.48\pm 0.05$\\
 GRB Number & 15 &19 &11 &8 & 53 \\
\hline
\\
$V-L$&$a=51.53\pm 0.27$&$a=52.25\pm 0.24$ & $a=52.38\pm 0.48$& $a=53.03\pm 0.42$& $a=51.86\pm 0.15$ \\
 &$b=0.50\pm 0.44$ &$b=0.35\pm 0.26$ & $b=0.25\pm 0.40$ & $b=-0.23\pm 0.33$  & $b=0.59\pm 0.15$  \\
   $\sigma_{\rm int}$ & $0.78\pm 0.13$ & $0.62\pm 0.10$ & $0.64\pm 0.15$ & $0.50\pm 0.11$ & $0.68\pm 0.06$ \\
 GRB Number & 26 &25 &15 &15 & 81 \\
\hline
\\
$\epkk-L$&$a=51.75\pm 0.12$&$a=52.21\pm 0.10$ & $a=52.27\pm 0.09$& $a=52.49\pm 0.10$& $a=52.11\pm 0.05$ \\
  &$b=1.35\pm 0.23$ &$b=1.29\pm 0.21$ & $b=1.40\pm 0.24$ & $b=0.55\pm 0.26$  & $b=1.40 \pm 0.12$  \\
   $\sigma_{\rm int}$ & $0.63\pm 0.09$ & $0.48\pm 0.08$ & $0.40\pm 0.07$ & $0.43\pm 0.07$ & $0.54\pm 0.04$\\
 GRB Number & 34 &30 &27 &25 & 116 \\
\hline
\\
$\epkk-E_\gamma$&$a=50.59\pm 0.10$&$a=50.66\pm0.11$ & $\cdots$ & $\cdots$ & $a=50.60\pm 0.07$ \\
 &$b=1.54\pm 0.21$ &$b=1.55\pm 0.34$ & $\cdots$ & $\cdots$ & $b=1.47\pm 0.20$  \\
   $\sigma_{\rm int}$ & $0.24\pm 0.11$ & $0.14\pm 0.13$ & $\cdots$ & $\cdots$ & $0.31\pm 0.08$\\
 GRB Number & 10 &7 &4 &3& 24 \\
\hline
\\
$\trt-L$&$a=52.49\pm 0.18$&$a=52.81\pm 0.11$ & $a=52.56\pm 0.18$& $a=52.86\pm 0.16$& $a=52.68\pm 0.07$ \\
  &$b=-1.20\pm 0.26$ &$b=-0.77\pm 0.21$ & $b=-1.03\pm 0.54$ & $b=-0.96\pm 0.57$  & $b=-1.12\pm 0.14$  \\
   $\sigma_{\rm int}$ & $0.50\pm 0.08$ & $0.39\pm 0.07$ & $0.55\pm 0.16$ & $0.43\pm 0.14$ & $0.48\pm 0.05$\\
 GRB Number & 25 &22 &13 &12& 72 \\
\hline
\\
$\epkk-E_{\rm \gamma, iso}$&$a=52.48\pm 0.13$&$a=52.84\pm 0.11$ & $a=52.74\pm 0.08$& $a=52.92\pm 0.10$& $a=52.71\pm 0.05$ \\
  &$b=1.47\pm 0.25$ &$b=1.54\pm 0.26$ & $b=1.29\pm 0.21$ & $b=0.89\pm 0.25$  & $b=1.47\pm 0.12$  \\
   $\sigma_{\rm int}$ & $0.62\pm 0.10$ & $0.54\pm 0.09$ & $0.36\pm 0.07$ & $0.39\pm 0.07$& $0.49\pm 0.04$ \\
 GRB Number & 28 &26 &25 &22& 101 \\
\hline
\\
\end{tabular}
\caption{\label{tablebs2007} Results of fits to the luminosity
  correlations for GRBs in each redshift bin and the whole sample.}
\end{table*}

\begin{table*}
  \centering
  \begin{tabular}{ccccccc}
    \hline
    & $\tlag-L$ & $V-L$ & $\epkk-L$ & $\epkk-E_\gamma$ & $\trt-L$ &
    $\epkk-E_{\rm \gamma, iso}$
    \\
    \hline
    $\mathrm{d}a / \mathrm{d}z$ & $0.22 \pm 0.08$ & $0.44 \pm 0.15$ & $0.18 \pm 0.04$
    & $\cdots$ & $0.06 \pm 0.07$ & $0.09 \pm 0.04$
    \\
    $\mathrm{d}b / \mathrm{d}z$ & $-0.06 \pm 0.12$ & $-0.23 \pm 0.15$ & $-0.22 \pm 0.10$
    & $\cdots$ & $0.07 \pm 0.18$ & $-0.19 \pm 0.10$
    \\
    \hline
  \end{tabular}
  \caption{The slopes of the parameters $a$ and $b$ versus redshift.}
  \label{tab:ab_slope}
\end{table*}

\section{Constraints on cosmological parameters and dark energy equation of state}

To constrain the cosmological parameters, we simultaneously fit
correlation parameters of GRBs and cosmological parameters to avoid
the circularity problem. Since the luminosity correlations of
$E_{\mathrm{peak}}-E_{\gamma}$ and $E_{\mathrm{peak}}-E_{\gamma,
\mathrm{iso}}$ describe almost the same physics, we can only include
one of them in the fit to avoid strong correlation among the
luminosity correlations. We choose the
$E_{\mathrm{peak}}-E_{\gamma}$ correlation, which has a smaller
intrinsic scatter. Since the intrinsic scatter of the $V-L$
correlation has been too large, including it in the fit or not has
little effect on the result. For the flat $\Lambda$CDM model, the
combination of the correlations gives the result of
$\omm=0.31^{+0.13}_{-0.10}$. Schaefer (2007) used a combination of
the same correlations with a smaller sample of GRBs and got the result
of $\omm \simeq 0.39$. Our result is consistent with the value of
Schaefer (2007) at the $1\sigma$ confidence level.

We also constrain the dark energy EOS using the GRBs together with
SNe Ia and the $H(z)$ data. We adopt the redshift binned
parametrization for the dark energy EOS, as proposed in Huterer \&
Cooray (2005), in which the redshifts are divided into several bins
and the dark energy EOS is taken to be constant in each redshift bin
but can vary from bin to bin. For this parametrization,
$f(z)=\rho_{\rm DE}(z)/\rho_{\rm DE}(0)$ takes the form (Sullivan et
al. 2007)
\begin{equation}
  \label{eq:fzbinned}
  f(z_{n-1}<z \le z_n)=
  (1+z)^{3(1+w_n)}\prod_{i=0}^{n-1}(1+z_i)^{3(w_i-w_{i+1})},
\end{equation}
where $w_i$ is the EOS parameter in the $i^{\mathrm{th}}$ redshift
bin defined by an upper boundary at $z_i$, and the zeroth bin is
defined as $z_0=0$. Such a parametrization scheme assumes less about
the nature of the dark energy, especially at high redshift, compared
with other simple parametrizations, since independent parameters are
introduced in every redshift range and it could, in principle,
approach any functional form with the increase of the number of
redshift bins (of course, we would need enough observational data to
constrain all the parameters well). For a given set of observational
data, the parameters $w_i$ are usually correlated with each other,
i.e. the covariance matrix
\begin{equation}
  \textbf{C}
  =\langle \textbf{w} \textbf{w}^{\mathrm{T}} \rangle
  - \langle \textbf{w} \rangle
  \langle \textbf{w}^{\mathrm{T}} \rangle
  ,
\end{equation}
is not diagonal. A new set of dark energy EOS parameters
$\widetilde{w_i}$ defined by
\begin{equation}
  \label{eq:transformation}
  \widetilde{\textbf{w}}=\textbf{T} \textbf{w}
  .
\end{equation}
is introduced to diagonalize the covariance matrix. The
transformation of $\textbf{T}$ advocated by Huterer \& Cooray (2005)
has the advantage that the weights (rows of $\textbf{T}$) are
positive almost everywhere and localized in redshift fairly well, so
the uncorrelated EOS parameters $\widetilde{w_i}$ are easy to
interpret intuitively. The evolution of the dark energy with respect
to the redshift could be estimated from these decorrelated EOS
parameters. The transformation of $\textbf{T}$ is determined as
follows. First, we define the Fisher matrix
\begin{equation}
  \label{eq:fisher_matrix}
  \textbf{F}\equiv\textbf{C}^{-1}
  =\textbf{O}^{\mathrm{T}}\Lambda \textbf{O}
  ,
\end{equation}
and then the transformation matrix $\textbf{T}$ is given by
\begin{equation}
  \label{eq:transf_matrix1}
  \textbf{T}=\textbf{O}^{\mathrm{T}}
  \Lambda^{\frac{1}{2}}\textbf{O}
  ,
\end{equation}
except that the rows of the matrix $\textbf{T}$ are normalized such
that
\begin{equation}
  \label{eq:transf_matrix2}
  \sum_j T_{ij}=1
  .
\end{equation}
We divided redshifts at points $z = 0.2, 0.5, 1$ and Markov chain
Monte Carlo techniques are used with $O(10^6)$ samples generated for
each result. Since current observational data have only very weak
constraints on the nature of dark energy at redshifts $z > 1$ (we
tried constraining the dark energy EOS without imposing any prior on
$w(z>1)$ using the parameterization described above with GRBs and
other data sets, no substantial constraints on the dark energy EOS at
redshifts $z>1$ can be obtained), we simply set $w(z>1)=-1$, and focus
on the dark energy EOS at $z\leq 1$.

In addition to GRBs, we have used Union2 compilation of SNe Ia from
Amanullah et al. (2010), BAO measurement from Percival et al. (2010)
and $\Omega_m h = 0.213 \pm 0.023$ from Tegmark et al. (2004). We
assumed the prior $\Omega_k=-0.014 \pm 0.017$ (Spergel et al. 2007)
for the cosmic curvature. We also used the $H(z)$ data from Stern et
al. (2010) and Riess et al. (2009).

For each luminosity correlation for GRBs, the $\chi_{\mathrm{GRB}}^2$
is calculated by
\begin{equation}
  \label{eq:chi2_GRB}
  \chi_{\mathrm{GRB}}^2 = -2 \ln L
  ,
\end{equation}
where $L$ is given by Eq.~(\ref{eq:likelihood}) except that
cosmological parameters are free parameters now. For other data set as
well as the priors, the usual way of calculating $\chi^2$ is used,
i.e., for a physical quantity $\xi$ with experimentally measured value
$\xi_o$, standard deviation $\sigma_{\xi}$, and theoretically
predicted value $\xi_t(\theta)$, where $\theta$ is a collection of
parameters needed to calculate the theoretical value, the $\chi^2$
value is given by
\begin{equation}
  \label{eq:chi2_xi}
  \chi_{\xi}^2(\theta)=\frac{
    \left(
      \xi_t(\theta)-\xi_o
    \right)^2
  }{\sigma_{\xi}^2}
  .
\end{equation}
The total $\chi_{\rm total}^2$ is the sum of all the $\chi^2$s from
independent data.

\begin{figure}
  \includegraphics[width = 0.45 \textwidth]{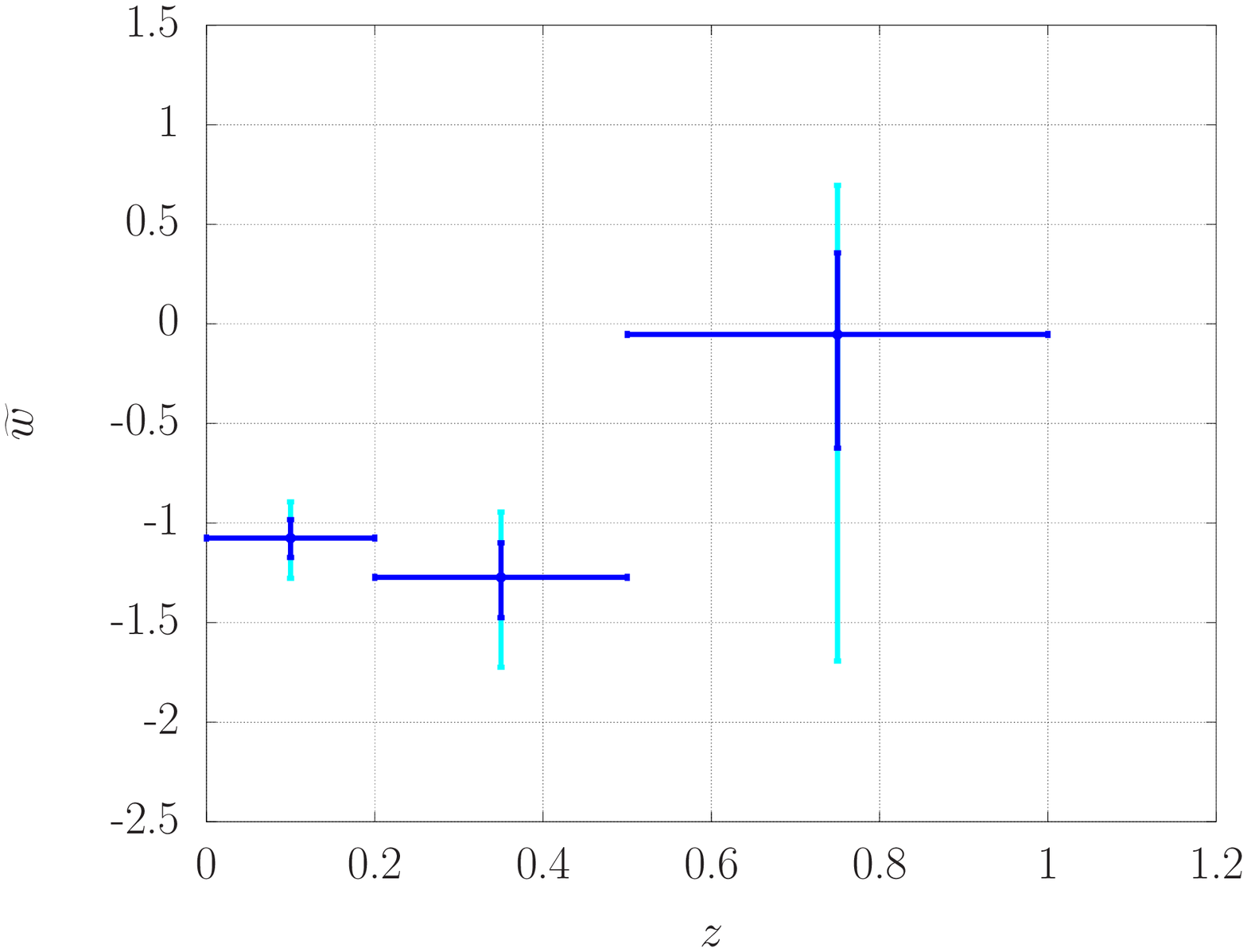} \\
  \includegraphics[width = 0.45 \textwidth]{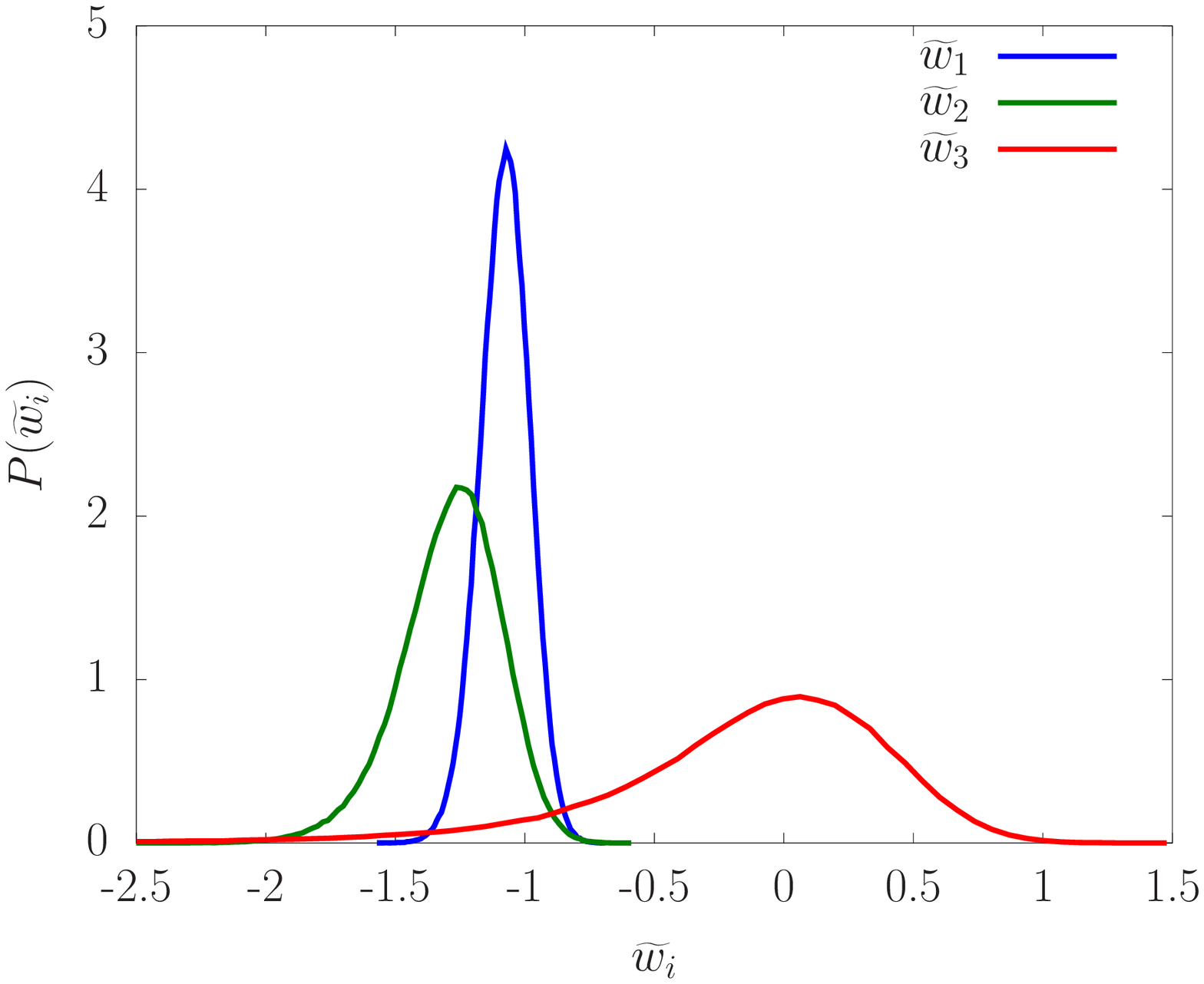} \\
  \includegraphics[width = 0.45 \textwidth]{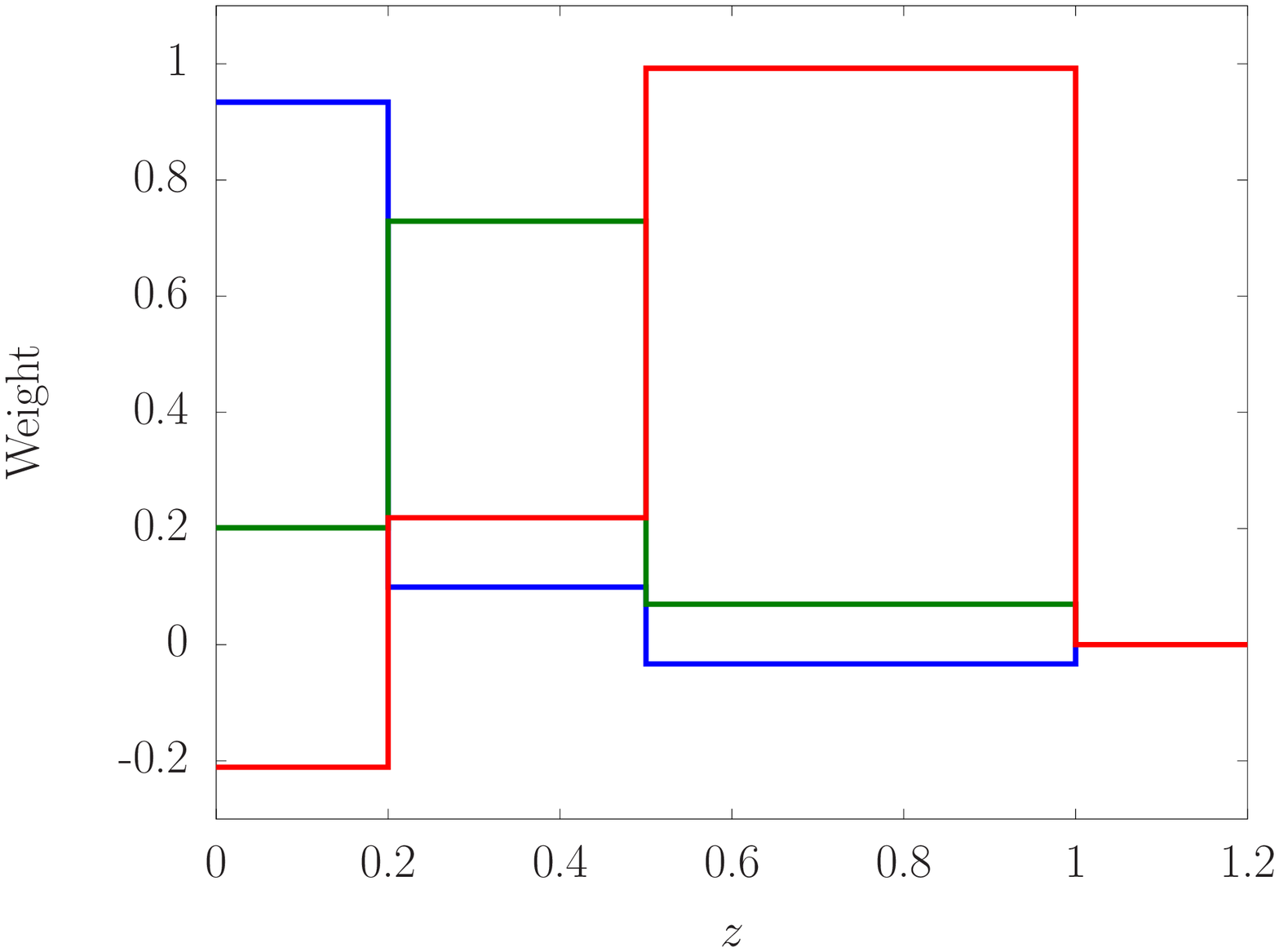}
  \caption
  {
    Estimates of the uncorrelated dark energy EOS parameters
    $\widetilde{w}_i$.
    Top panel: uncorrelated dark energy parameters versus redshift, in which
    the vertical errorbars correspond to $1 \sigma$ and $2 \sigma$
    confidence levels of $\widetilde{w}_i$ and the horizontal
    errorbars span the corresponding redshift bins from which the
    contributions to $\widetilde{w}_i$ come most.
    Middle panel: Probability distribution for $\widetilde{w}_i$.
    Bottom panel: window functions for $\widetilde{w}_i$.
  }
  \label{fig:wz_and_weight}
\end{figure}

\begin{figure}
  \includegraphics[width = 0.45 \textwidth]{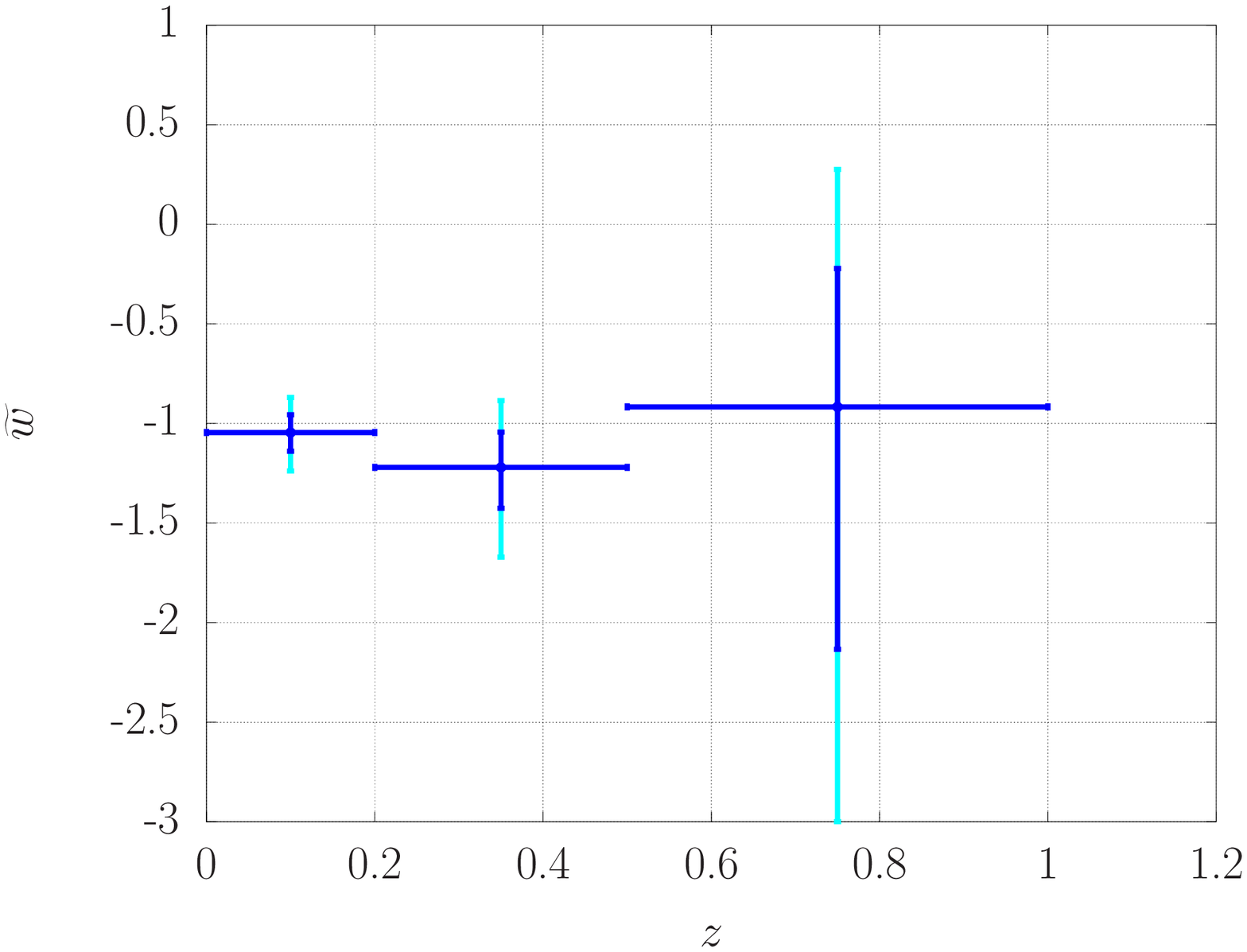} \\
  \includegraphics[width = 0.45 \textwidth]{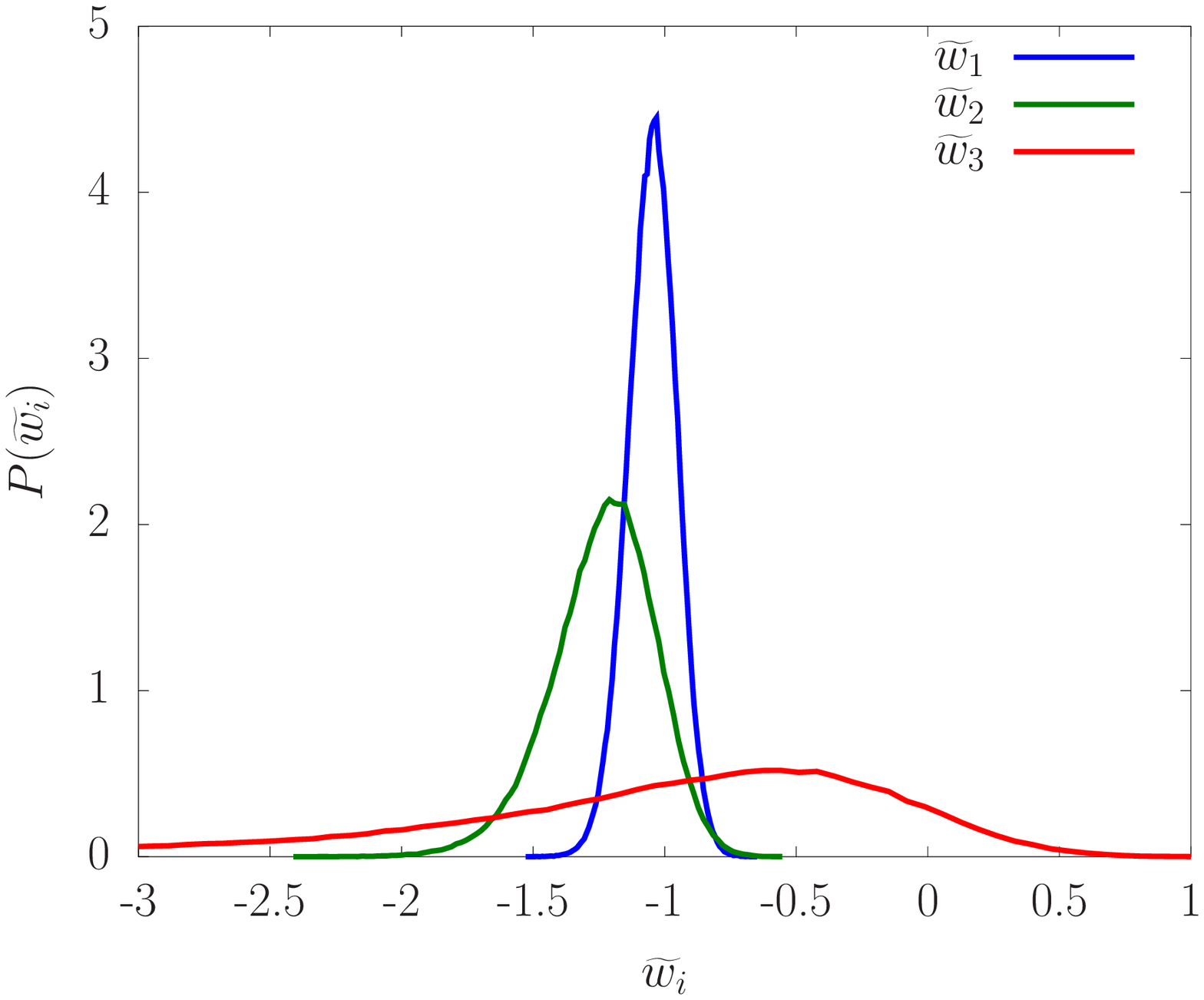} \\
  \includegraphics[width = 0.45 \textwidth]{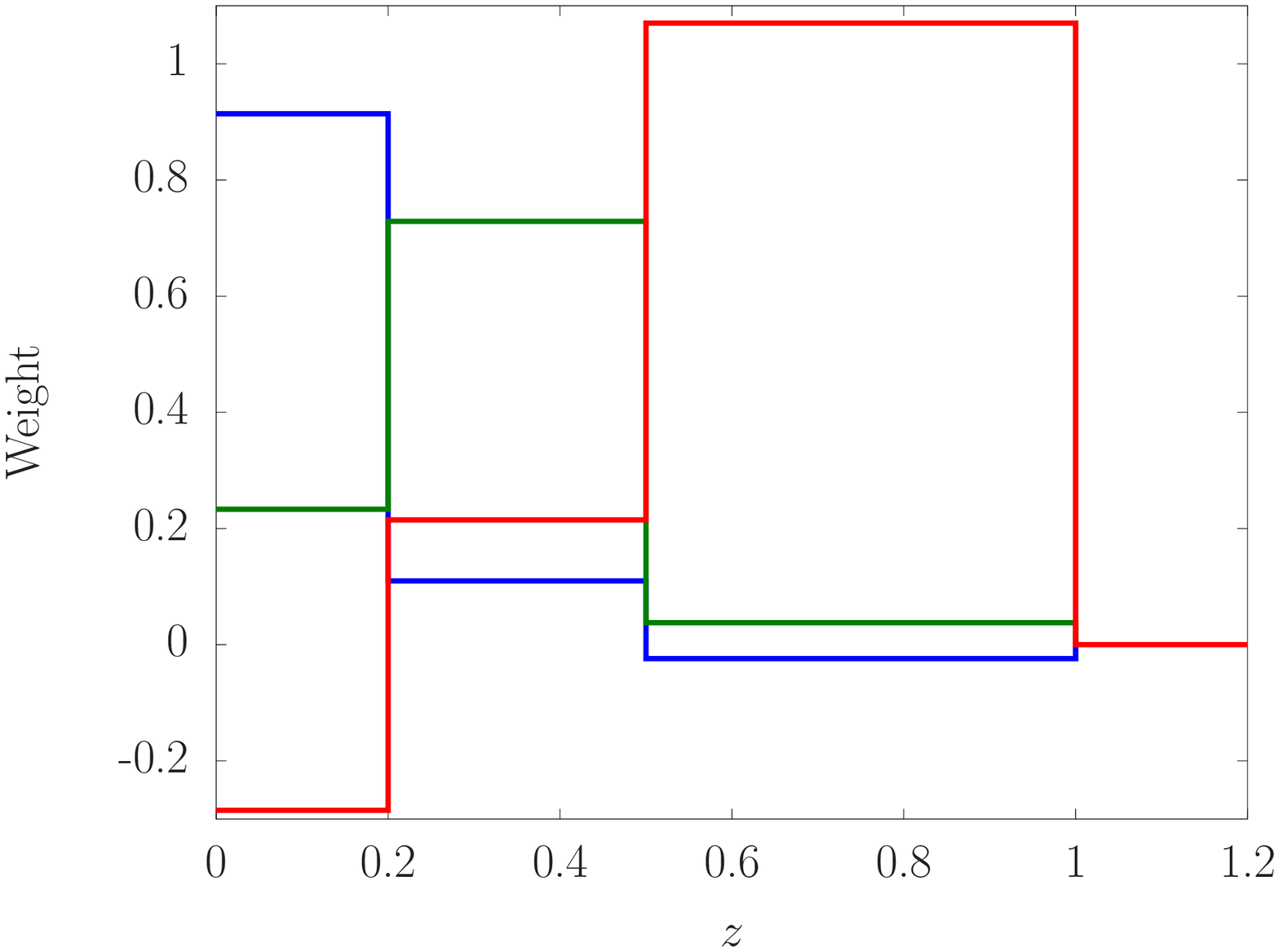}
  \caption
  {
    Estimates of the uncorrelated dark energy EOS parameters
    $\widetilde{w}_i$. Same as the Figure~\ref{fig:wz_and_weight}
    except that GRBs are not included in the fit.
  }
  \label{fig:wz_and_weightnoGRB}
\end{figure}

Figure~\ref{fig:wz_and_weight} shows the result derived from the
data set described above. We can see that though the dark energy is
consistent with the cosmological constant ($w(z)=-1$) at the $2
\sigma$ confidence level, there is still considerable room for an
evolving dark energy EOS. Notably, the slight deviation of the dark
energy from the cosmological constant at $z \geq 0.5$, which
persistently appears with many previous data sets, still exists
here. For our result here, the deviation is mainly due to the GRBs.
Though the Union compilation of SNe Ia gives the same trend of
deviation from the cosmological constant due to the unexpected
brightness of the Hubble data at $z>1$, when the sample is enlarged,
it seems that such a character of the SN Ia data has been averaged
out in the Union2 compilation. As a comparison, we also present in
Figure~\ref{fig:wz_and_weightnoGRB} the result derived from the data
set without GRBs included. See also Wang et al. (2010) and Park et
al. (2010) for similar analysis on the nature of the dark energy
with Union2 compilation of SNe Ia. The deviation of the dark energy
EOS from $-1$ may arise from many possible reasons, for example, the
statistical errors due to the limitation of current observational
data, some biasing systematic errors in the observational data
(especially there is still some distance to calibrating GRBs as
ideal standard candles), or the nature of the dark energy itself,
etc. It should be made clear about the cause of the deviation with
future observational data in order to understand the dark energy
better.

\section{Conclusions}

To build up the Hubble diagram to a redshift higher than the one of
SNe Ia, most attempts have been devoted to search for a method to
make GRBs standardizable candles. Different correlations have been
proposed in order to build up a GRB Hubble diagram and constrain
cosmological parameters. As a further step, we have here considered
the latest GRB dataset and luminosity correlations to constrain the
cosmological parameters and dark energy.

In this paper, we derived the six luminosity correlations
($\tlag-L$, $V-L$, $\epkk-L$, $\epkk-E_\gamma$, $\trt-L$,
$\epkk-E_{\gamma, \mathrm{iso}}$) from the light curves and spectra of
the latest 116
long GRBs. We find that the intrinsic scatter of $V-L$ correlation
is too large and there seems no inherent correlation between the two
parameters using the latest GRB data. The other five correlations
indeed exist when enlarging the sample. We have found no
statistically significant evidence for the redshift evolution of
the luminosity correlations. However, even the best GRB luminosity
correlation is currently not competitive with other
cosmological probes of the cosmic acceleration expansion since the
cosmological parameter $1\sigma$ errors derived from GRBs
($\omm=0.31^{+0.13}_{-0.10}$) are more than an order of magnitude
larger than the corresponding errors obtained using SN Ia standard
candles and other geometrical probes. But the estimates of
cosmological parameters from GRBs are important because they provide
an independent confirmation of the results from other probes.

We also performed an investigation on the dark energy EOS using the
GRBs together with the Union2 compilation of SNe Ia and the $H(z)$
data. The result is consistent with the cosmological constant at $2
\sigma$ confidence level. However, mainly due to the GRB data, the
slight deviation of the dark energy EOS from $-1$ at $z \geq 0.5$,
which persistently appears with many previous data sets, still
exists.

\section*{Acknowledgements}
This work is supported by the National Natural Science Foundation of
China (grants 10873009 and 11033002) and the National Basic Research
Program of China (973 program) No. 2007CB815404 (for ZGD). SQ is
supported by the National Natural Science Foundation of China under
grant No. 10973039, the Jiangsu Planned Projects for Postdoctoral
Research Funds under grant No. 0901059C and the China Postdoctoral
Science Foundation under grant No. 20100471421. FYW is supported by
Jiangsu Planned Projects for Postdoctoral Research Funds 1002006B
and China Postdoctoral Science Foundation funded project
20100481117.

\end{document}